\documentclass[twocolumn]{aastex631}

\usepackage[caption = false]{subfig}
\usepackage{csquotes}

\usepackage[normalem]{ulem}
\usepackage{xcolor}

\begin{document}

\title{Generation of sub-ion scale magnetic holes from electron shear flow instabilities in plasma turbulence}

\author{Giuseppe Arrò}
\affiliation{Department of Mathematics, KU Leuven, Leuven, Belgium}
\affiliation{Dipartimento di Fisica \enquote{E. Fermi}, Università di Pisa, Pisa, Italy}

\author{Francesco Pucci}
\affiliation{Istituto per la Scienza e Tecnologia dei Plasmi, Area della Ricerca di Bari, Bari, Italy}

\author{Francesco Califano}
\affiliation{Dipartimento di Fisica \enquote{E. Fermi}, Università di Pisa, Pisa, Italy}

\author{Maria Elena Innocenti}
\affiliation{Institut f\"ur Theoretische Physik, Ruhr-Universit\"at Bochum, Bochum, Germany}

\author{Giovanni Lapenta}
\affiliation{Department of Mathematics, KU Leuven, Leuven, Belgium}

\begin{abstract}

Magnetic holes (MHs) are coherent structures associated with strong magnetic field depressions in magnetized plasmas. They are observed in many astrophysical environments at a wide range of scales but their origin is still under debate. In this work we investigate the formation of sub-ion scale MHs using a fully kinetic 2D simulation of plasma turbulence initialized with parameters typical of the Earth's magnetosheath. Our analysis shows that the turbulence is capable of generating sub-ion scale MHs from large scale fluctuations via the following mechanism: first, the nonlinear large scale dynamics spontaneously leads to the development of thin and elongated electron velocity shears; these structures then become unstable to the electron Kelvin-Helmholtz instability and break up into small scale electron vortices; the electric current carried by these vortices locally reduces the magnetic field, inducing the formation of sub-ion scale MHs. The MHs thus produced exhibit features consistent with satellite observations and with previous numerical studies. We finally discuss the kinetic properties of the observed sub-ion scale MHs, showing that they are characterized by complex non-Maxwellian electron velocity distributions exhibiting anisotropic and agyrotropic features. 

\end{abstract}


\section{Introduction} 

Magnetic holes (MHs) are coherent structures characterized by a significant and sharp reduction in the magnetic field amplitude. They have been observed and studied for several years in a variety of astrophysical environments. The earliest observations of MHs date back to the work of \citet{turner1977magnetic}, where these structures were identified for the first time in the solar wind (SW). Since then, satellite measurements have revealed that MHs are abundant not only in the SW \citep{winterhalter1995magnetic,winterhalter2000latitudinal,chisham2000multisatellite,russell2008mirror,zhang2008characteristic,wang2020statistical,xiao2014plasma,wang2021foreshock,wang2021first}, but also in the Earth's magnetotail \citep{ge2011case,sun2012cluster,balikhin2012magnetic,huang2019mms,shustov2019statistical,liu2021kinetics}, planetary magnetosheaths \citep{johnson1997global,soucek2008properties,volwerk2008mirror,yao2017observations,huang2017magnetospheric,zhong2019observations,yao2020propagating,karlsson2021magnetic,huang2021situ,goodrich2021evidence}, planetary bow shocks \citep{cattaneo1998evolution,wang2021statistical,chen2022electron,huang2022kinetic} and around comets \citep{russell1987mirror,volwerk2008mirror}. 
Magnetic holes come in many different sizes, covering a very broad range of scales. The largest MHs span up to hundreds of ion gyroradii $\rho_i$ \citep{turner1977magnetic,stevens2007scale,zhang2008characteristic,ahmadi2018generation}, while the smallest ones are sub-ion scale structures whose size can be a few electron gyroradii $\rho_e$ \citep{sundberg2015properties,goodrich2016electric,gershman2016electron,zhang2017kinetics,yao2017observations}. 

Many observational \citep{winterhalter1994ulysses,erdHos1996statistical,soucek2008properties,zhang2008characteristic,xiao2014plasma} and numerical studies \citep{genot2009mirror,ahmadi2017simulation} seem to support the idea that large scale MHs may emerge from the nonlinear evolution of mirror modes. However, the mirror instability usually tends to generate magnetic peaks \citep{hellinger2017mirror}, while holes develop only under very specific conditions \citep{baumgartel2003towards,califano2008nonlinear,kuznetsov2015variational}. Therefore, the origin of large scale MHs is still under debate.

Sub-ion scale MHs have received more attention in recent years thanks to the advent of new space missions, such as the Magnetospheric Multiscale (MMS) mission \citep{burch2016magnetospheric}, that made it possible to probe kinetic scales with great accuracy. Differently from the large scale ones, sub-ion scale MHs display properties that are inconsistent with the hypothesis of them being generated by the mirror instability \citep{sundberg2015properties}. In particular, their size is much smaller than the typical wavelengths associated with mirror modes, and the environment they are observed in is often stable to the mirror instability \citep{balikhin2012magnetic}. Furthermore, these sub-ion scale structures are typically associated with an enhanced electron temperature anisotropy, while the ions do not show any response to the presence of such small scale MHs. This is in contrast with the properties of large scale MHs, where the magnetic field depression is balanced by an increase in ion density and pressure. All these discrepancies have brought to the conclusion that large scale and sub-ion scale MHs are essentially different structures generated by different mechanisms. 

Several processes have been considered as possible mechanisms for the generation of sub-ion scale MHs. Since these structures are often characterized by a perpendicular electron temperature anisotropy, it has been argued that they may emerge from the nonlinear development of the electron mirror and field-swelling instabilities \citep{basu1982field,basu1984theory,marchenko1988drift,gary2006linear,pokhotelov2013physical,hellinger2018electron,liu2021kinetics}, representing the extension of the large scale mirror instability to sub-ion scales in the case of hot electrons with temperature $T_e$ larger than the ion temperature $T_i$ \citep{migliuolo1986field}. However, the condition $T_e\!>\!T_i$ is rarely satisfied where sub-ion scale MHs are observed \citep{liu2021kinetics}, so the electron mirror and field-swelling instabilities often fail to provide a proper explanation for the generation of these structures \citep{balikhin2012magnetic,sundberg2015properties}. In \citet{balikhin2012magnetic}, the authors argue that the electron tearing instability may be responsible for generating sub-ion scale magnetic field depressions that later evolve into MHs. This scenario seems suitable to explain the observations of \citet{ge2011case} but is unlikely in many other cases \citep{sundberg2015properties}. Sub-ion scale MHs have also been described in terms of electron magnetohydrodynamics (EMHD) solitons \citep{ji2014emhd,li2016emhd}, whose properties compare well with observations \citep{yao2016propagation}. Therefore, also in the specific case of sub-ion scale MHs, there is still no general agreement about their generation mechanism.

Plasma turbulence is known to drive the formation of coherent structures \citep{karimabadi2013coherent,wan2016intermittency} and may play a fundamental role in setting up favorable conditions for the generation of sub-ion scale MHs. Indeed, sub-ion scale MHs have been observed in the turbulent magnetosheath \citep{huang2016mms,huang2017magnetospheric,huang2017statistical} and in numerical simulations of freely decaying plasma turbulence \citep{haynes2015electron,roytershteyn2015generation}. \citet{haynes2015electron} showed that sub-ion scale MHs can develop self-consistently in 2D plasma turbulence from small scale magnetic field fluctuations that locally reduce the magnetic field intensity. Such fluctuations can trap electrons, generating a ring-shaped current consistent with the magnetic field depression and a perpendicular electron temperature anisotropy that sustains the structure, making it stable for more than 100 electron gyroperiods. Due to the presence of the azimuthal electron current, these structures were dubbed electron vortex magnetic holes (EVMHs). \citet{roytershteyn2015generation} showed that EVMHs may also develop in 3D simulations of turbulence and their properties are consistent with many observations of sub-ion scale MHs \citep{sundberg2015properties,goodrich2016electric,gershman2016electron,goodrich2016mms,zhang2017kinetics,huang2017magnetospheric,yao2017observations,zhong2019observations,wang2020study}. However, in both \citet{haynes2015electron} and \citet{roytershteyn2015generation}, the authors observe that the EVMHs develop at relatively early stages of their simulations and likely emerge from the relaxation of the initial small scale magnetic perturbations introduced to trigger the turbulence. This leaves open the question of whether the formation of these EVMHs is actually driven by the turbulence or induced by the small scale energy injection in the initial condition. Moreover, in turbulence the energy is typically injected at large scales and cascades to small scales due to nonlinear interactions among perturbations. The nature of sub-ion scale fluctuations is thus influenced by the large scale dynamics that determines the local plasma properties, including those that can favor the formation of sub-ion scale MHs. Hence, a key problem is understanding how large scale turbulent fluctuations can dynamically set up the conditions for the formation of sub-ion scale MHs.

In this work, we investigate the formation of sub-ion scale EVMHs in a fully kinetic particle-in-cell (PIC) simulation of 2D freely decaying plasma turbulence. We consider a plasma with properties similar to those typically observed in the Earth's magnetosheath and we initialize the turbulence with large scale perturbations whose wavelengths span several ion inertial lengths $d_i$. We show that in our simulation the turbulence naturally leads to the formation of sub-ion scale electron velocity shears associated with sheet-like magnetic field depressions whose length can reach up to about $10\,d_i$. These elongated velocity shears then become unstable to the electron Kelvin-Helmholtz instability (EKHI) and break up into sub-ion scale electron vortices associated with localized magnetic dips that can evolve into EVMHs. This dynamics represents one of the possible mechanisms capable of producing sub-ion scale MHs from large scale fluctuations in plasma turbulence. The observed EVMHs have features consistent with previous numerical simulations and satellite observations. We finally analyze the kinetic properties of a number of EVMHs generated by the aforementioned mechanism, showing that these structures are characterized by non-trivial anisotropic and agyrotropic electron velocity distributions exhibiting different features depending on local plasma conditions.

\section{Simulation setup}

We performed a fully kinetic simulation of freely decaying plasma turbulence using the semi-implicit energy conserving PIC code ECsim \citep{markidis2011energy,lapenta2017exactly,lapenta2023advances}. The simulation domain is represented by a 2D square periodic box of size $L\!=\!64\,d_i$, sampled by an uniform mesh containing $2048^2$ points. We consider an ion-electron plasma with a reduced ion-to-electron mass ratio of $m_i/m_e\!=\!100$. Particles are initialized from a Maxwellian distribution and we employ $5000$ particles per cell for both ions and electrons. The plasma is initially quasi-neutral with uniform density. The initial temperature is uniform and isotropic, with plasma beta equal to $\beta_i\!=\!8$ for the ions and to $\beta_e\!=\!2$ for the electrons. The electron inertial length is $d_e\!=\!0.1\,d_i$ while the ion and electron gyroradii are initially equal to $\rho_i\!=\!\sqrt{\beta_i}\,d_i\!\simeq\!2.83\,d_i$ and $\rho_e\!=\!\sqrt{\beta_e}\,d_e\!\simeq\!1.41\,d_e$, respectively. The initial magnetic field configuration includes an uniform out-of-plane guide field $\textbf{B}_0\!=\!B_0\,\hat{\textbf{z}}$ (with $\hat{\textbf{z}}$ being the unit vector in the out-of-plane direction) and the turbulence is triggered by adding random phase, isotropic magnetic field and velocity fluctuations to the uniform background. The wavenumbers $k$ of the initial perturbations fall in the range $1\leqslant k/k_0 \leqslant 4$ (where $k_0\!=\!2\pi/L$), corresponding to an injection scale of about $\lambda_{inj}\!\simeq\!16\,d_i$. The root mean square (rms) amplitude of magnetic field fluctuations $\delta \textbf{B}$ is $\delta B_{rms}/B_0\!\simeq\!0.9$ while the rms amplitude of ion and electron fluid velocity fluctuations $\delta \textbf{u}$ is $\delta u_{rms}/c_A\!\simeq\!3.6$ (with $c_A$ being the initial Alfvén speed based on the guide field $B_0$). This initialization aims at reproducing conditions similar to those typically observed in the Earth's magnetosheath \citep{phan2018electron,stawarz2019properties,bandyopadhyay2020situ}. The ratio between the electron cyclotron frequency $\Omega_e$ and the ion cyclotron frequency $\Omega_i$ is $\Omega_e/\Omega_i\!=\!m_i/m_e\!=\!100$ while the ratio between the plasma frequency and the cyclotron frequency is $\omega_{p,i}/\Omega_i\!=\!100$ for the ions and $\omega_{p,e}/\Omega_e\!=\!10$ for the electrons. The time step used in the simulation is equal to $\Delta t\!=\!0.05\, \Omega_e^{-1}$. Additional information about the simulation can be found in \citet{arro2022spectral}, where the spectral properties of the turbulence have been analyzed. 

\section{Results}

\subsection{Electron vortex magnetic holes at fully developed turbulence}

\begin{figure*}[t]
\centering
\subfloat{
\includegraphics[width=.95\linewidth]{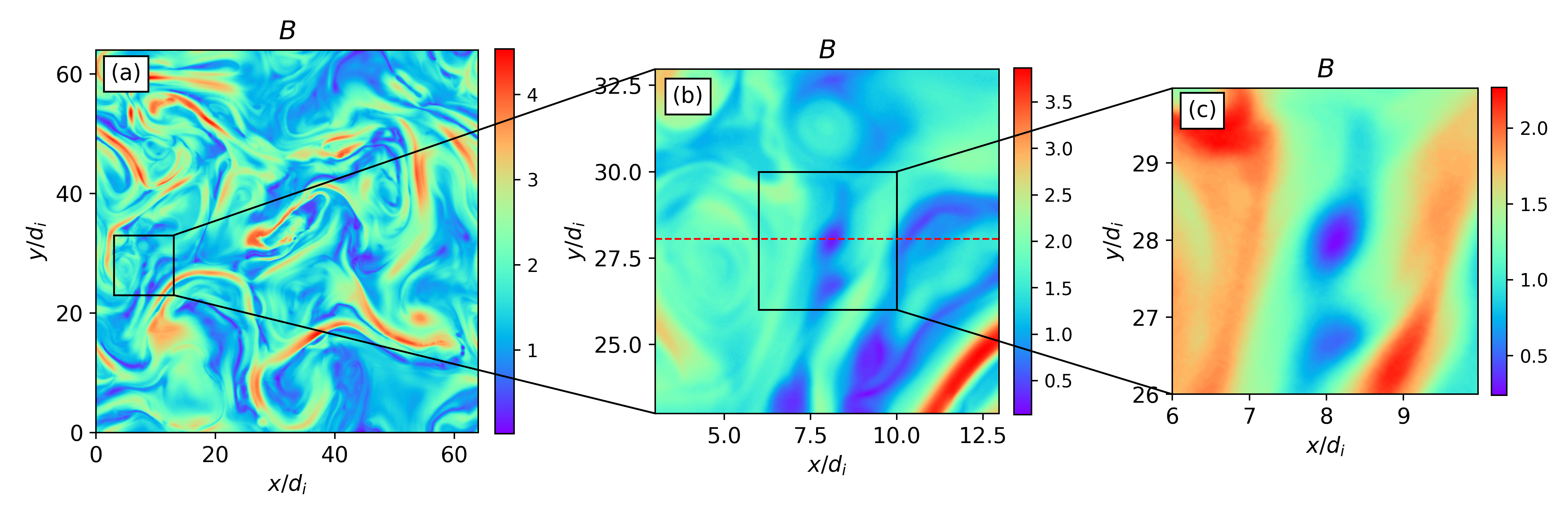} 
}
\vspace{-0.7cm}
\subfloat{
\includegraphics[width=.9\linewidth]{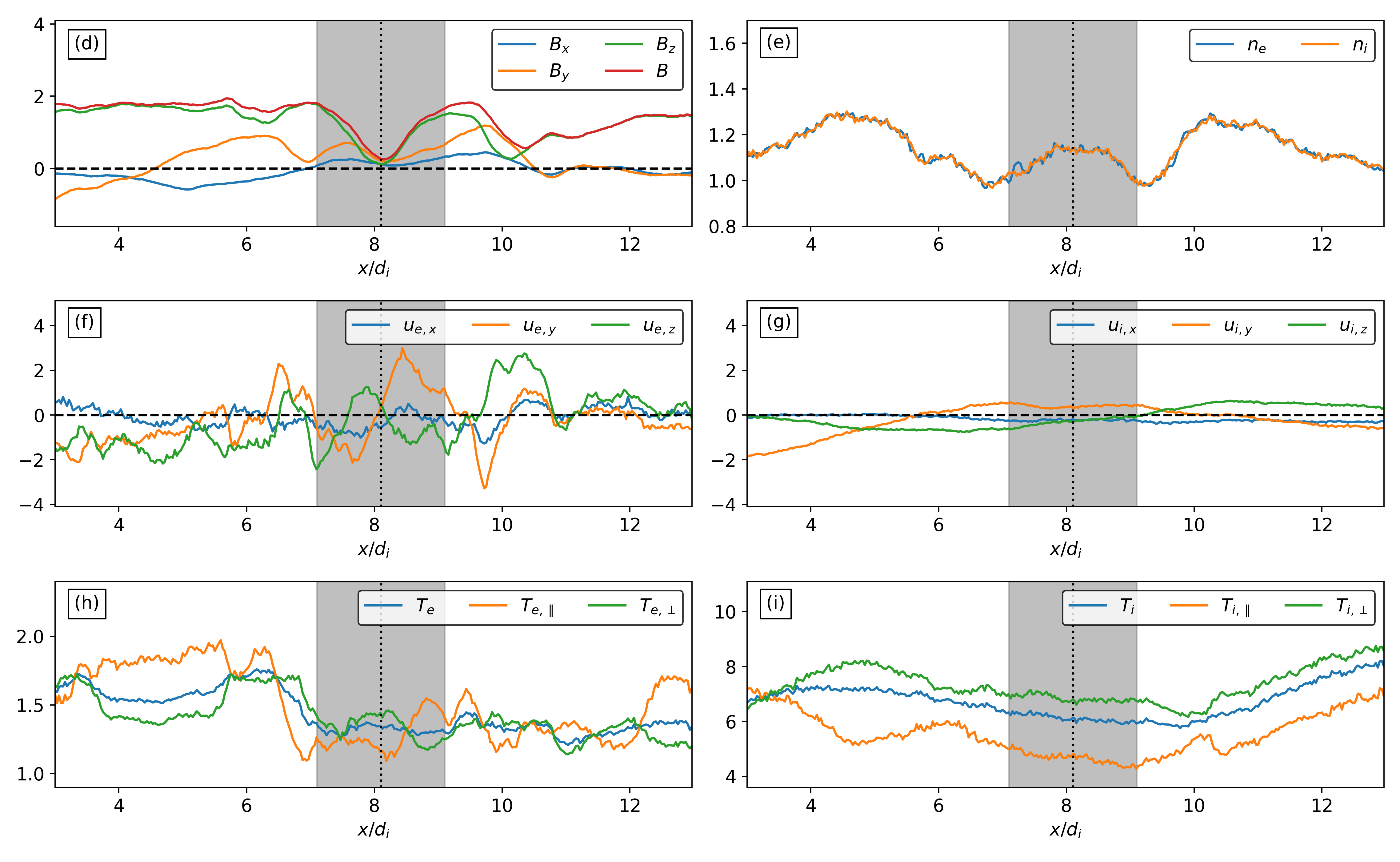} 
}
\caption{Panels (a)-(c): shaded isocontours of the magnetic field magnitude (in units of $B_0$) over the whole simulation box at $t\!=\!650\,\Omega_e^{-1}$, with a progressive zoom into an EVMH. Panels (d)-(i): Magnetic field (in units of $B_0$), electron and ion number densities (in units of the initial density), electron and ion fluid velocities (in units of the initial Alfvén speed $c_A$), electron and ion temperatures (in units of the initial electron temperature) over a 1D cut crossing the EVMH, indicated by the red dashed line in panel (b); black horizontal dashed lines are located at zero, the grey shaded area highlights a region of width $2\,d_i$ around the EVMH, whose center is indicated by the black vertical dotted line.}
\label{satellite}
\end{figure*}

Panel (a) of Figure~\ref{satellite} shows a 2D visualization of the full simulation domain at $t\!=\!650\,\Omega_e^{-1}$, when the turbulence is fully developed. The shaded isocontours represent the magnetic field magnitude and we observe a wide variety of magnetic structures at different scales. Panels (b) and (c) show a progressive close-up into a region where an oval-shaped sub-ion scale magnetic field depression is found. This structure is indeed an EVMH. Panels (d) to (i) of Figure~\ref{satellite} show different quantities plotted over a 1D cut crossing the EVMH, indicated by the red dashed line in panel (b) of the same figure. In panel (d) we see that the magnetic field intensity suddenly drops in correspondence of the EVMH. The magnetic depression mainly results from a reduction in $B_z$, which is the dominant component of the magnetic field. Panel (e) highlights a density increase associated with the magnetic dip, a feature typically observed in EVMHs. The density increase is caused by the fact that EVMHs tend to trap particles. The plasma is quasi-neutral in correspondence of the structure since the ion and electron densities are essentially equal. Panel (f) shows that all the components of the electron fluid velocity change sign across the center of the EVMH, indicating a bipolar flow in correspondence of the structure. The dominant component is $u_{e,y}$, which is perpendicular to the 1D cut. These signatures imply the presence of an electron vortex associated with the magnetic field dip. On the other hand, in panel (g) we see that the ion fluid velocity is smaller than the electron velocity and does not show any variation correlated with the presence of the EVMH. Finally, electron and ion temperatures are shown in panels (h) and (i), respectively. The two panels include the temperature components parallel and perpendicular to the local magnetic field. We see that the perpendicular electron temperature $T_{e,\perp}$ increases in correspondence of the EVMH while the parallel component $T_{e,\parallel}$ decreases. This electron temperature anisotropy characterized by $T_{e,\perp}\!>\!T_{e,\parallel}$ is another typical property of EVMHs. As for the ions, no evident temperature variation is observed in connection with the EVMH. The fact that the ions do not show any response to the magnetic dip is consistent with them being demagnetized at scales of the order of the EVMH size. All the properties of the EVMH we showed are consistent with previous numerical simulations and satellite observations \citep[e.g.][]{haynes2015electron,huang2017statistical}.

As seen in Figure~\ref{satellite}, the diameter of these EVMHs is of the order of $d_i$, much smaller than both the domain size and the injection scale. Hence, the EVMHs are definitely produced by the turbulence. Indeed, a significant number of EVMHs are observed in our simulation at fully developed turbulence. We see that the turbulent motion dynamically produces and sometimes destroys these structures that therefore are an important and abundant element of plasma turbulence. Our numerical approach allows for the study of EVMHs formation from the beginning of the simulation to fully developed turbulence. This study is the focus of the next section.

\subsection{Electron vortex magnetic holes formation}

\begin{figure*}[t]
\centering
\includegraphics[width=.98\linewidth]{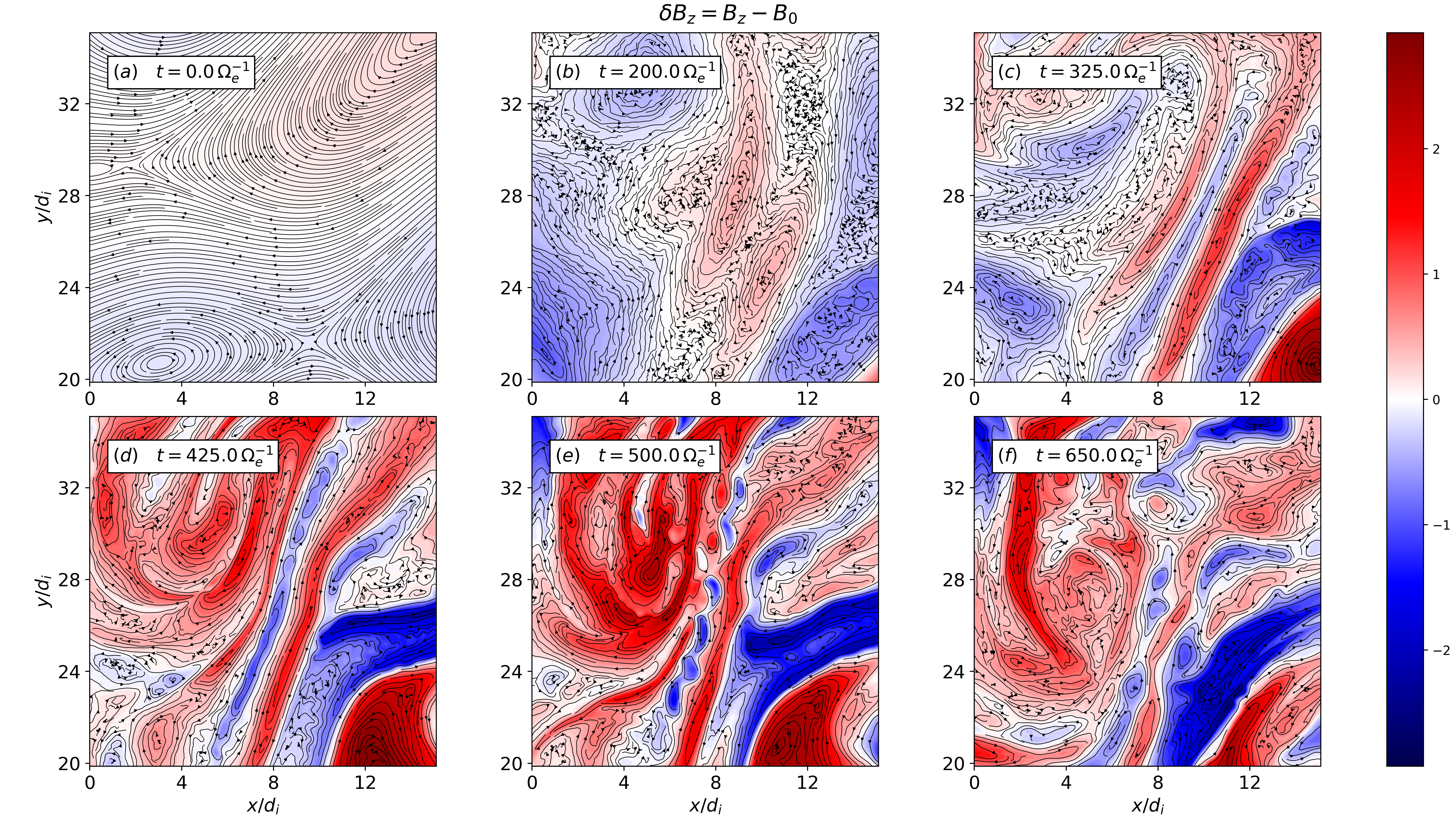} 
\caption{Time sequence describing the formation of a sub-ion scale magnetic hole from large scale turbulent fluctuations. Shaded isocontours represent the out-of-plane magnetic field fluctuations $\delta B_z$ (in units of $B_0$) while black streamlines indicate the in-plane electron fluid velocity in the frame comoving with the local center of mass velocity $\textbf{u}^{\prime}_e$.}
\label{Formation}
\end{figure*}

In this section we discuss the formation of sub-ion scale MHs driven by large scale turbulent fluctuations. 

Figure~\ref{Formation} shows a close-up of a region in our simulation where an EVMH forms, at different time steps. The shaded isocontours represent the $z$ component of magnetic field fluctuations, defined as $\delta B_z\!=\!B_z\!-\!B_0$ (where $B_0$ is the initial guide field intensity), while the black streamlines indicate the in-plane electron fluid velocity in the frame comoving with the local center of mass velocity $\textbf{u}$, i.e. $\textbf{u}^{\prime}_e\!=\!\textbf{u}_e-\textbf{u}$ (where $\textbf{u}_e$ is the electron fluid velocity and the superscript $\cdot^\prime$ indicates the frame comoving with $\textbf{u}$). This change of frame is performed to separate the small scale electron motion from the underlying large scale collective bulk flow of ions and electrons. The magnetic field is not affected by this change of frame since for small velocities with respect to the speed of light $c$, corrections to the magnetic field intensity are of second order in $\|\textbf{u}\|/c$ and thus negligible \citep{biskamp1997nonlinear}. Starting from panel (a) of Figure~\ref{Formation}, we see that initially only large scale fluctuations are present in the system, introduced by the initialization consisting of perturbations with wavelengths $\lambda$ in the range $16\,d_i\!\leqslant\!\lambda\!\leqslant\!64\,d_i$. As they evolve over time, these large amplitude fluctuations gradually change their shape and size due to the nonlinear dynamics, as observed in panel (b) of Figure~\ref{Formation} where the initial perturbations are being stretched and deformed while growing in amplitude. This nonlinear turbulent dynamics eventually leads to the formation of an elongated electron velocity shear associated with a magnetic field depression, as observed in panel (c) of Figure~\ref{Formation} where this structure is visible in the center of the figure as a sheet-like magnetic field dip characterized by $\delta B_z\!<\!0$ (squashed between two stretched regions with $\delta B_z\!>\!0$), with oppositely directed electron velocity streamlines at its edges. Initially, the electron velocity shear layer has a length that spans about $10\,d_i$ and a width smaller than $d_i$. The presence of a magnetic field depression in correspondence of the electron velocity shear is a consequence of the fact that at the scale of this structure only the electrons are coupled to the magnetic field dynamics while the ions are demagnetized. This means that there are no small scale ion velocity perturbations developing in response to the formation of the magnetic dip and the velocity shear is sustained solely by the electrons. This implies the presence of a net small scale in-plane current density $\textbf{J}_\perp$ carried exclusively by the electrons that balances the out-of-plane magnetic field variations, as consistent with the Ampere's law $\nabla_{\perp} \!\times\! \left(B_z\,\hat{\textbf{z}}\right)\!\sim\!\textbf{J}_\perp$. After its formation, the elongated electron velocity shear maintains its longitudinal structure which remains roughly stable (except for a slight large scale bending about its center) while its width gradually shrinks, indicating a steepening of the velocity shear and the consequent deepening of the magnetic field depression. Because of the underlying non-homogeneous turbulent motion, the shrinking is not uniform along the structure and its lower part shrinks slightly faster than the upper part. The steepening proceeds for about $100\,\Omega_e^{-1}$, until the width of the velocity shear has reached a scale of the order of $d_e$. At that point the structure becomes unstable, as observed in panel (d) of Figure~\ref{Formation} where small amplitude sinusoidal electron velocity fluctuations with wavelength of the order of $d_i$ have developed around the perimeter of the magnetic dip, perturbing its structure. The amplitude of these fluctuations grows over time until the instability enters the nonlinear stage characterized by the formation of large amplitude vortices that progressively tear apart the initial elongated velocity shear, compromising its structure. This is shown in panel (e) of Figure~\ref{Formation} where a chain of tilted electron velocity vortices has developed, breaking apart the electron velocity shear. Once the instability growth has saturated, after about $75\,\Omega_e^{-1}$, the velocity shear is finally dismantled and the vortices start to evolve and drift away from their initial position, advected by the underlying large scale flow. Some of these vortices are later destroyed as a consequence of the interaction among themselves or with other turbulent structures, but others survive for a sufficiently long time and eventually evolve into EVMHs. This can be observed in panel (f) of Figure~\ref{Formation} where some of the vortices previously generated by the instability have disappeared, swallowed by the surrounding turbulent structures, while one of them has evolved into an EVMH, centered at position $(x, y)\!\simeq\!(8.1\,d_i, 28\,d_i)$. The diameter of the EVMH is of the order of $d_i$, consistently with the wavelength of the unstable modes from which it has developed.

\begin{figure*}[t]
\centering
\subfloat{
\includegraphics[width=.48\linewidth]{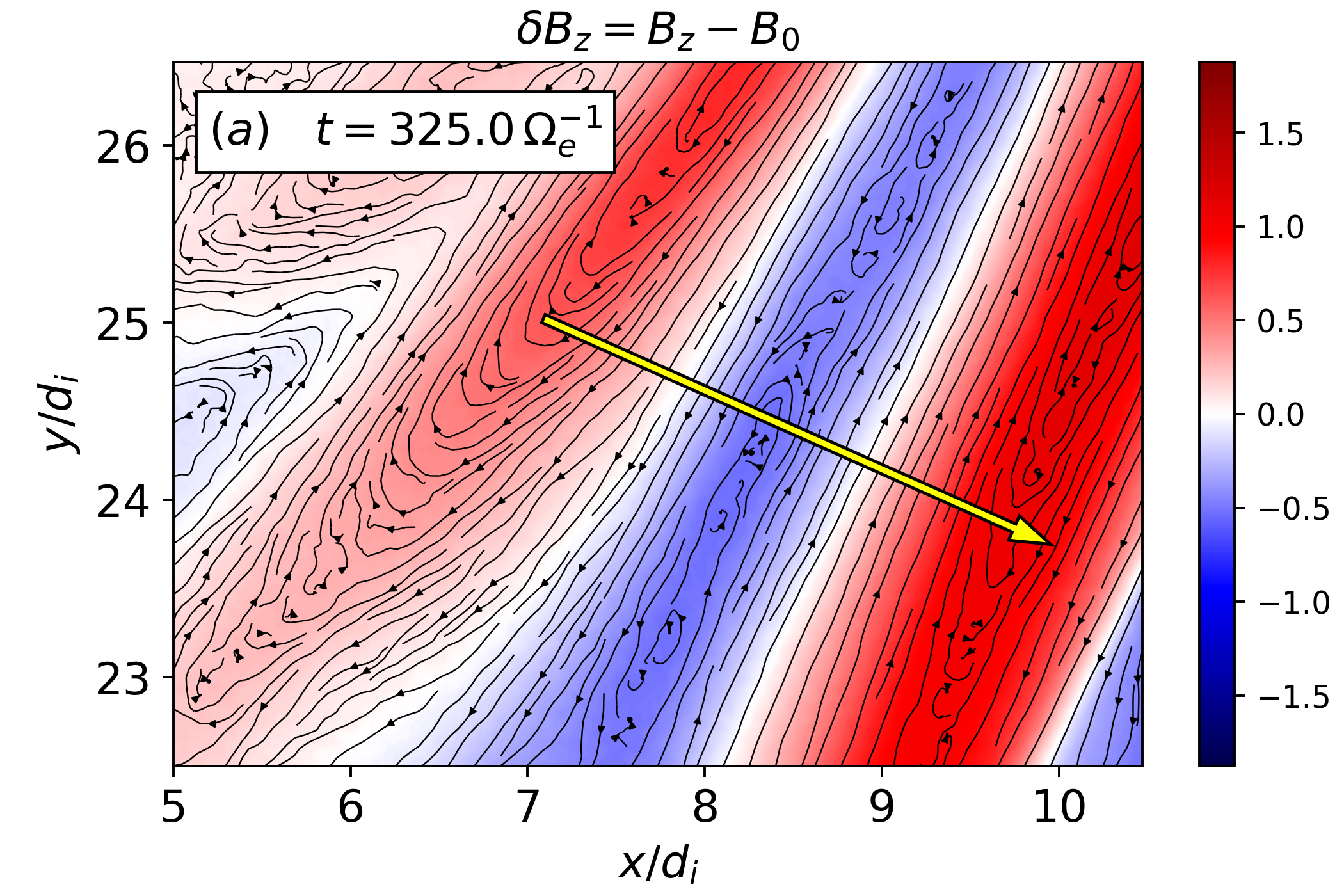} 
}
\hfill
\subfloat{
\includegraphics[width=.48\linewidth]{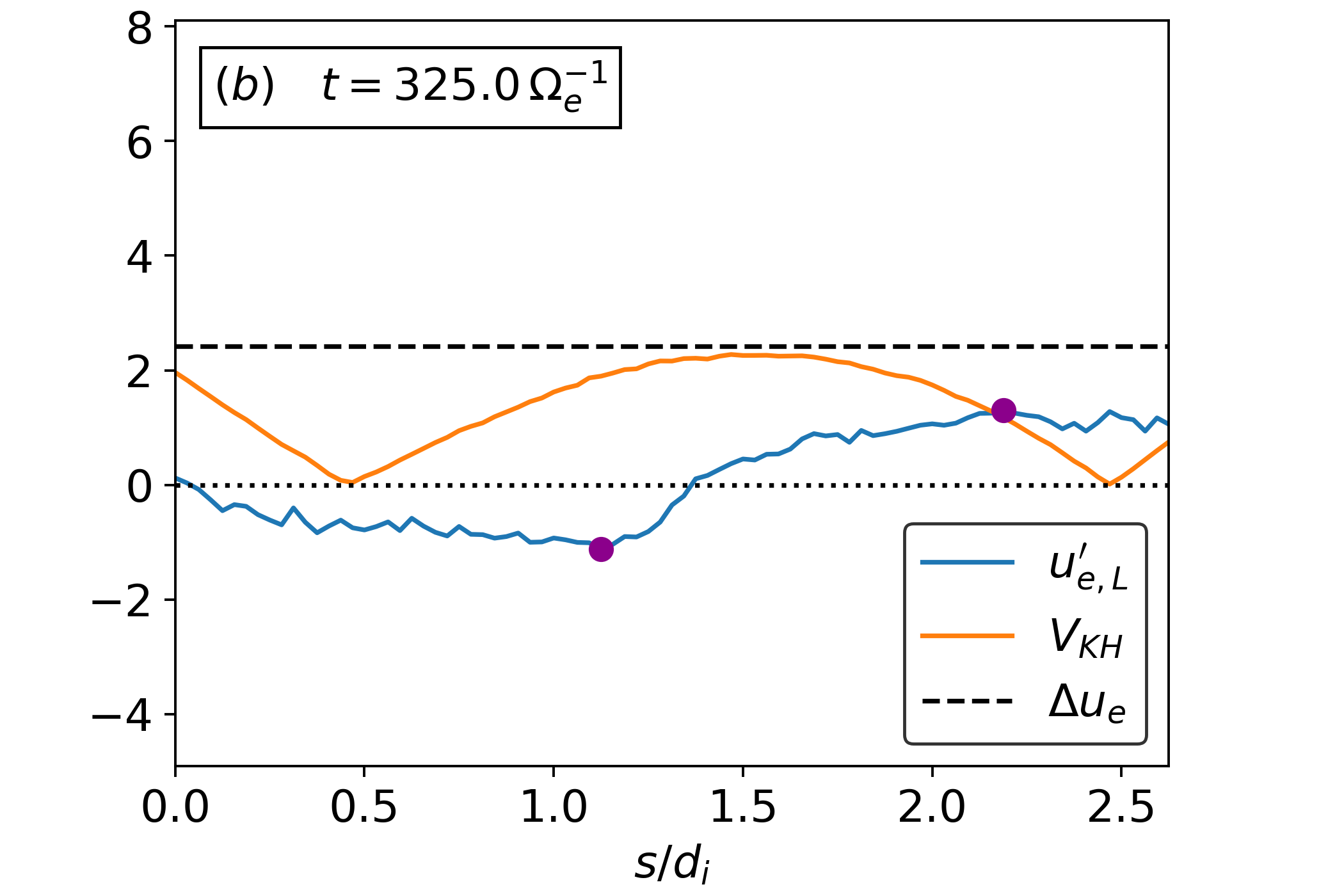}
}
\\
\subfloat{
\includegraphics[width=.48\linewidth]{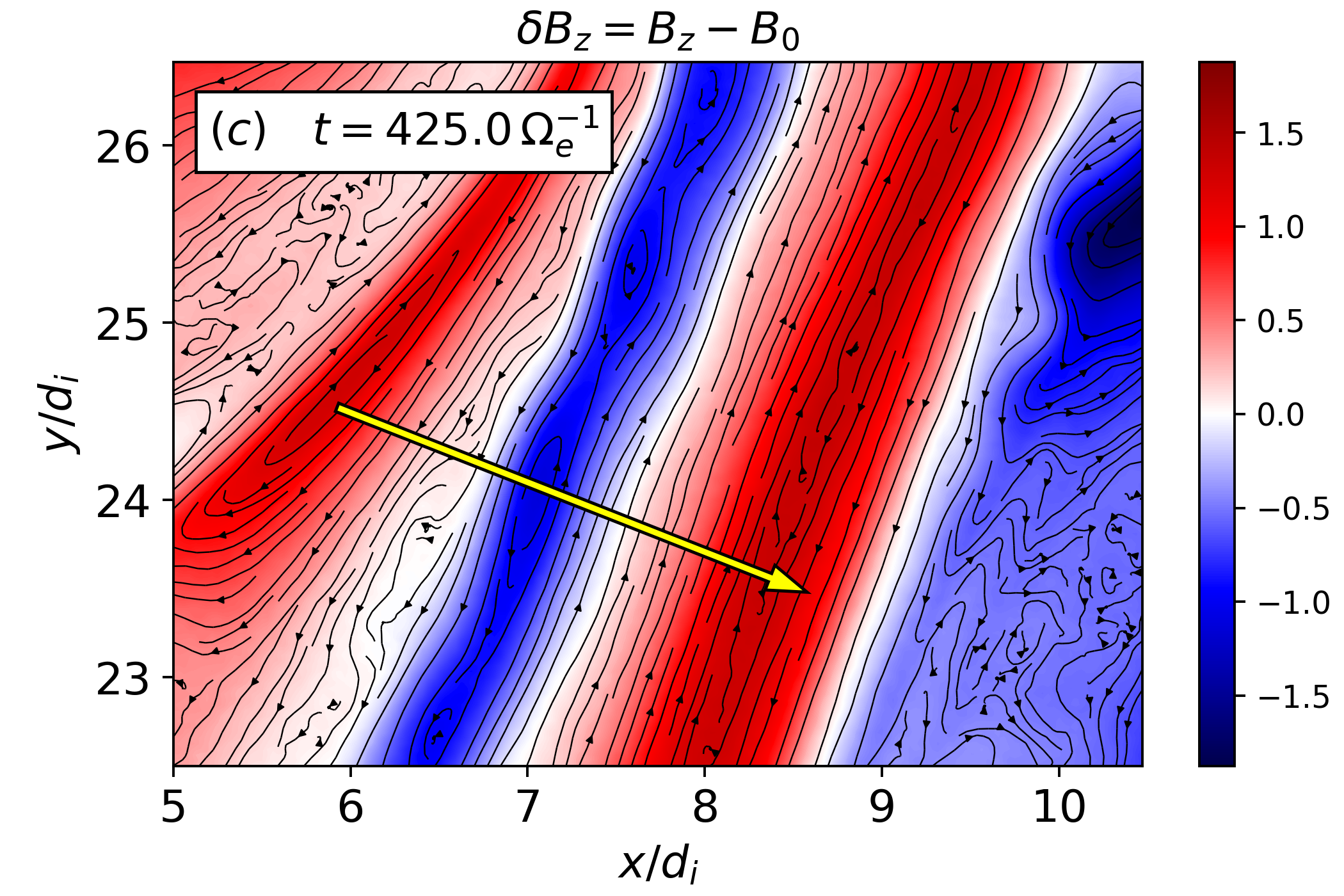} 
}
\hfill
\subfloat{
\includegraphics[width=.48\linewidth]{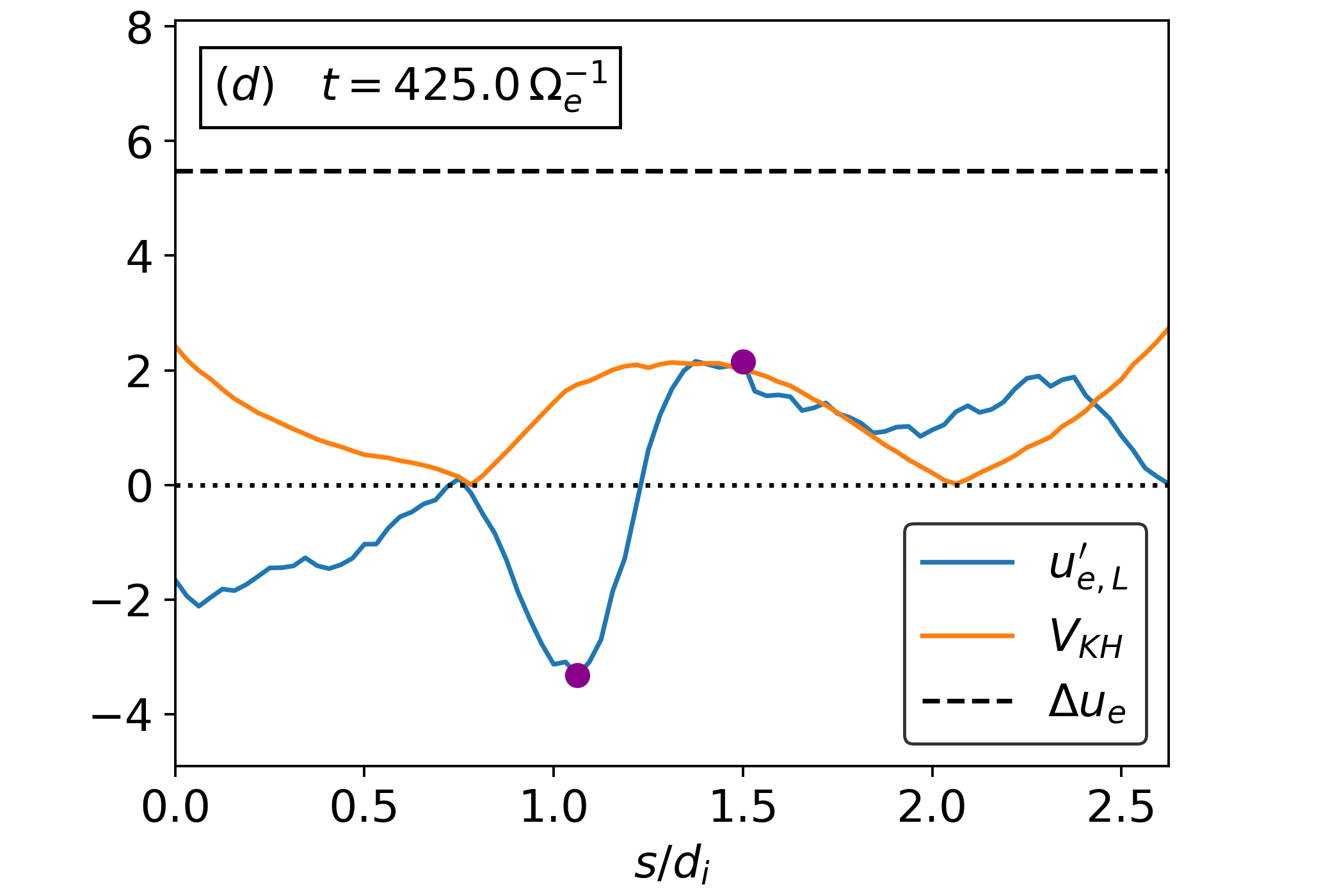}
}
\caption{Left panels: close-up of the electron velocity shear at $t\!=\!325\,\Omega_e^{-1}$ (a) and $t\!=\!425\,\Omega_e^{-1}$ (c), where shaded isocontours indicate $\delta B_z$ (in units of $B_0$), black streamlines represent the in-plane electron fluid velocity $\textbf{u}^{\prime}_e$ and the yellow solid arrow indicates the direction of the transversal cut through the velocity shear. Right panels: electron fluid velocity longitudinal to the shear flow $u_{e,L}^{\prime}$, threshold velocity $V_{KH}$ and electron velocity jump $\Delta u_e$ (between the two purple dots) over the transversal cut at $t\!=\!325\,\Omega_e^{-1}$ (b) and $t\!=\!425\,\Omega_e^{-1}$ (d) (all velocities are in units of the initial Alfvén speed $c_A$ and the horizontal dotted line is located at zero).}
\label{Cut}
\end{figure*}

The example we have just discussed, illustrated in Figure~\ref{Formation}, is representative of the typical dynamics that leads to the formation of EVMHs in our simulation. Similar situations are observed in other regions of the system where the same dynamics is at play, following an analogous development. Although the details of the whole process are different from case to case and depend on local plasma conditions, some basic ingredients are always present, namely the formation of an elongated electron velocity shear and the way it becomes unstable, breaking apart into electron vortices that may eventually become EVMHs. The dynamics responsible for the disruption of the electron velocity shear, represented in Figure~\ref{Formation} and observed also in other cases of EVMHs formation, is qualitatively consistent with the development of an EKHI. This instability has been studied theoretically and numerically in the context of cold electron magnetohydrodynamics (EMHD) \citep{jain2003nonlinear,jain2004kink,das2003sausage,gaur2009role,gaur2012linear} and it has been previously observed both in fully kinetic simulations and satellite observations of magnetic reconnection, where it is responsible for the disruption of electron velocity shears developing in the diffusion region and in reconnection outflows \citep{pritchett2009asymmetric,fermo2012secondary,huang2015magnetic,zhong2018evidence,zhong2022stacked}. In \citet{gaur2009role} the EKHI has been investigated considering a velocity shear with an hyperbolic tangent profile, sustained by the electrons alone and therefore associated with a magnetic field variation perpendicular to the plane of the shear, similarly to what we observe in our simulation. The authors showed that the velocity shear becomes significantly unstable to the EKHI only when its width is of the order of $d_e$ or smaller, with a growth rate that is non-zero only for modes with wavelength larger than the shear width. It was also shown that in presence of a finite magnetic field parallel to the shear flow, the magnetic tension generated by unstable modes can excite the propagation of whistler waves that subtract energy from the instability, inducing a stabilizing effect that hinders the growth of the EKHI. As a consequence, in order for the EKHI to develop, its growth rate must exceed the whistler frequency, as discussed in \citet{fermo2012secondary} where an instability criterion based on this concept has been introduced. The EKHI growth rate $\gamma$ can be estimated as $\gamma\!\sim\!\Delta u_e/d_e$, where $\Delta u_e$ is the velocity jump across the shear whose width is of the order of $d_e$, while the frequency of whistler waves propagating parallel to the shear flow is $\omega\!=\!\Omega_{e,L}\,k^2\,d_e^2/(1+k^2\,d_e^2)$, where $\Omega_{e,L}$ is the electron cyclotron frequency based on the magnetic field component parallel to the shear flow. By imposing that $\gamma$ must be larger than the whistler frequency in order for the EKHI to develop, the instability criterion is obtained:\\
\begin{equation}
\Delta u_e > V_{KH}(k) := c_{A_e,L}\,\frac{k^2\,d_e^2}{1+k^2\,d_e^2}\,,
\label{Condition}
\end{equation}
\\where $c_{A_e,L}\!=\!d_e\,\Omega_{e,L}$ is the electron Alfvén speed parallel to the shear flow and $k$ is the wavenumber of the unstable mode. In the case of fluctuations with $k\!\sim\!1/d_e$, Equation~(\ref{Condition}) reduces to the instability condition $\Delta u_e\!>\!c_{A_e,L}/2$ used in \citet{fermo2012secondary}. 

To show that the instability we observe is also quantitatively consistent with an EKHI, we carry out a local stability analysis to characterize the properties of the electron velocity shear as it becomes unstable. For this analysis we consider and compare two times: $t\!=\!325\,\Omega_e^{-1}$, when the velocity shear has just formed and the instability has not developed yet, and $t\!=\!425\,\Omega_e^{-1}$, when the structure has reached its minimum width and is significantly unstable. In panel (a) of Figure~\ref{Cut} we show a close-up of the lower portion of the electron velocity shear at $t\!=\!325\,\Omega_e^{-1}$. The shaded isocontours indicate $\delta B_z$ and the black streamlines represent the in-plane electron fluid velocity $\textbf{u}^{\prime}_e$, with the same format as Figure~\ref{Formation}. As already discussed, at this point the velocity shear looks stable and perturbations have not significantly developed yet. To perform the stability analysis we consider a transversal cut through the electron velocity shear in the direction indicated by the yellow solid arrow. In panel (b) of Figure~\ref{Cut} we show the electron fluid velocity $\textbf{u}^{\prime}_e$ in the direction longitudinal to the shear flow, indicated as $u^{\prime}_{e,L}$, and the threshold velocity $V_{KH}$ defined in Equation~(\ref{Condition}), interpolated over the transversal cut. The direction longitudinal to the shear flow is defined by the unit vector $\hat{\textbf{l}}\!=\!\hat{\textbf{z}}\times\hat{\textbf{n}}$, where $\hat{\textbf{n}}$ is the unit vector representing the direction of the transversal cut (indicated by the yellow arrow in panel (a) of Figure~\ref{Cut}). The coordinate $s$ indicates the position on the transversal cut in the direction of $\hat{\textbf{n}}$. Since the plasma is not homogeneous in proximity of the electron velocity shear, $V_{KH}$ is calculated using the local magnetic field and density (most of the spatial variation comes from the longitudinal magnetic field). The wavenumber $k$ needed to calculate $V_{KH}$ is estimated using the typical wavelength of the unstable modes observed in panel (d) of Figure~\ref{Formation}, which is of the order of $d_i$, corresponding to $k\!\sim\!2\pi/d_i$. We also show an estimate of the velocity jump $\Delta u_e$ across the electron velocity shear, indicated by the black dashed horizontal line and calculated as the difference between the maximum and minimum values of the velocity $u^{\prime}_{e,L}$ in the interval considered, highlighted by the filled purple dots. As can be seen from panel (b) of Figure~\ref{Cut}, right after its formation the electron velocity shear is only slightly above the EKHI threshold, consistently with the absence of prominent unstable modes in panel (a) of the same figure. Panel (c) of Figure~\ref{Cut} shows a second close-up of the electron velocity shear at $t\!=\!425\,\Omega_e^{-1}$, where the plotted quantities are the same as in panel (a). As already mentioned, at this time the velocity shear has become narrower and appears significantly perturbed by the presence of sinusoidal modes with wavelength of the order of $d_i$. By looking at the transversal cut in panel (d) of Figure~\ref{Cut}, we see that the velocity shear has become steeper, with a width of the order of $d_e$, and the velocity jump has increased. The comparison between $\Delta u_e$ and $V_{KH}$ shows that the structure is now way above the instability threshold, consistently with the development of the EKHI.

The local stability analysis yields to the same conclusions also in the case of other electron velocity shears generated by the turbulence in our simulation (not shown here), showing that these structures are well above the EKHI instability threshold when unstable modes develop and start to grow, eventually leading to the velocity shear disruption and to the formation of EVMHs. Thus, our analysis indicates that the EKHI plays a crucial role in the turbulent cascade, acting as a bridge between large scales and sub-ion scales in the dynamics responsible for the formation of EVMHs. In other words, the EKHI represents a possible efficient mechanism capable of transferring energy from large scale turbulent fluctuations to sub-ion scale structures, inducing the development of EVMHs.

\subsection{Kinetic properties of electron vortex magnetic holes}

\begin{figure*}[t]
\centering
\begin{minipage}{0.32\linewidth}
\subfloat{
\includegraphics[width=\linewidth]{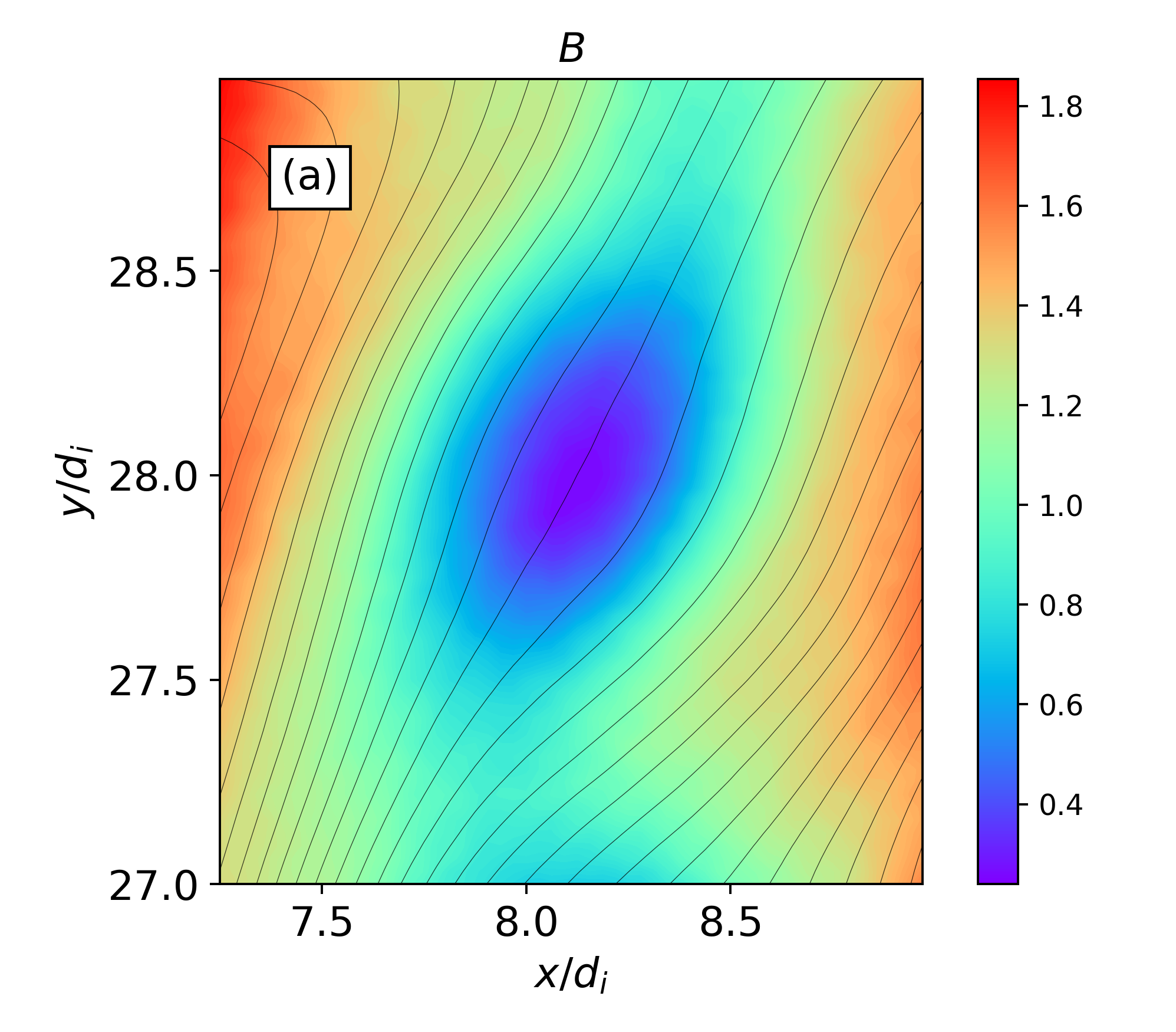} 
}
\vspace{-0.5cm}
\subfloat{
\includegraphics[width=\linewidth]{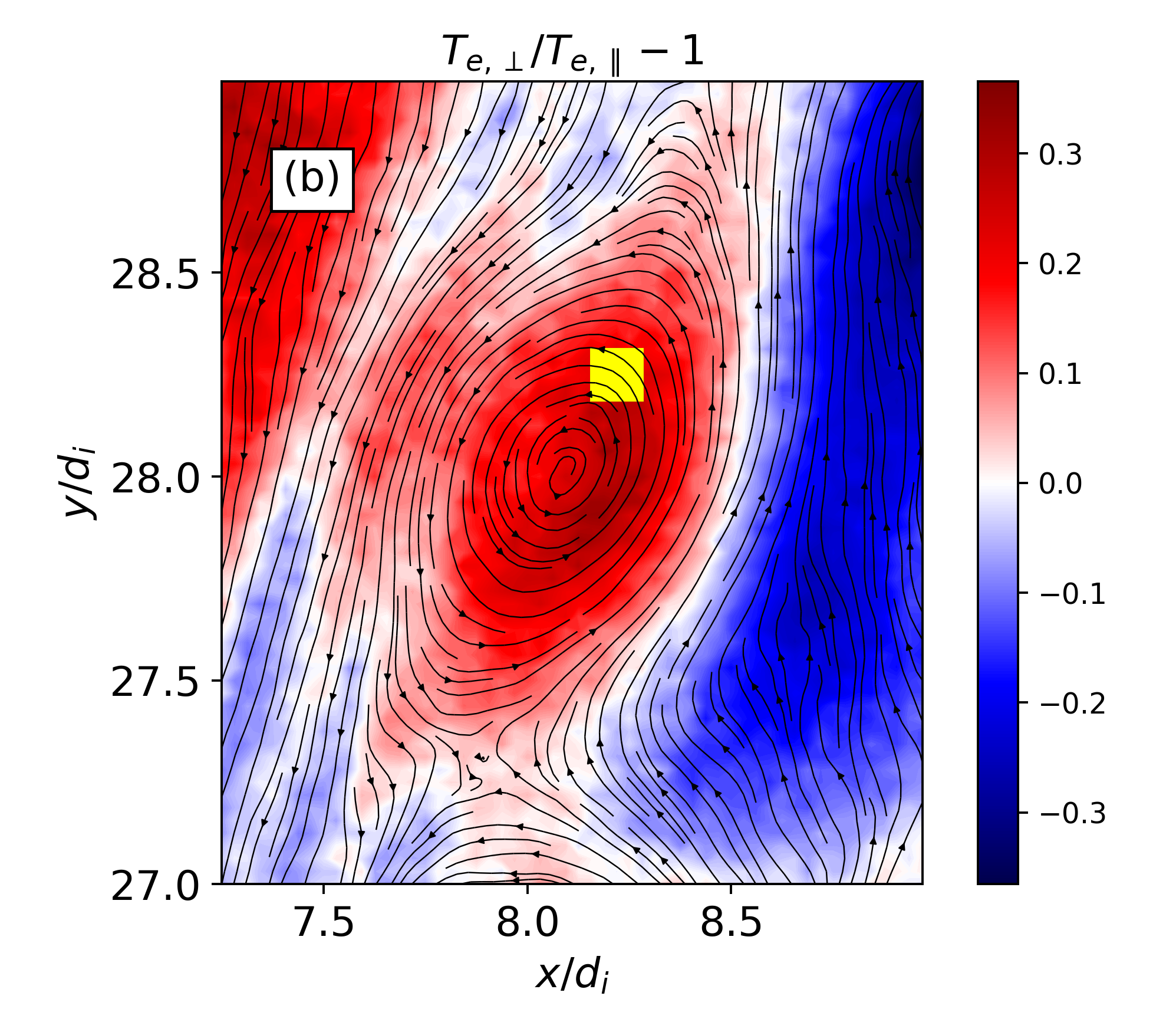} 
}
\end{minipage}
\hspace{0.9cm}
\begin{minipage}{0.58\linewidth}
\vspace{-0.5cm}
\subfloat{
\includegraphics[width=\linewidth]{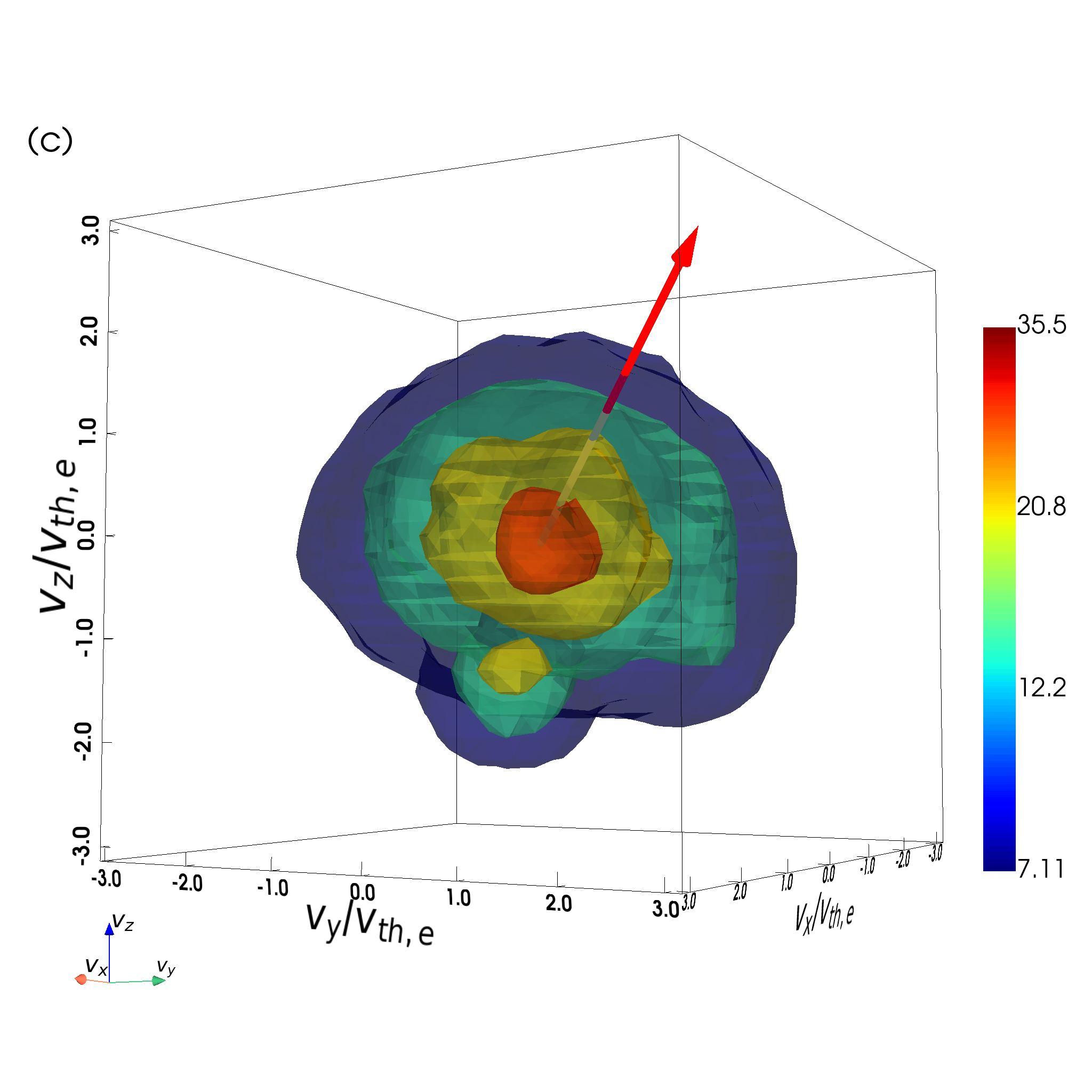}
}
\end{minipage}
\\
\begin{minipage}{\linewidth}
\subfloat{
\includegraphics[width=0.32\linewidth]{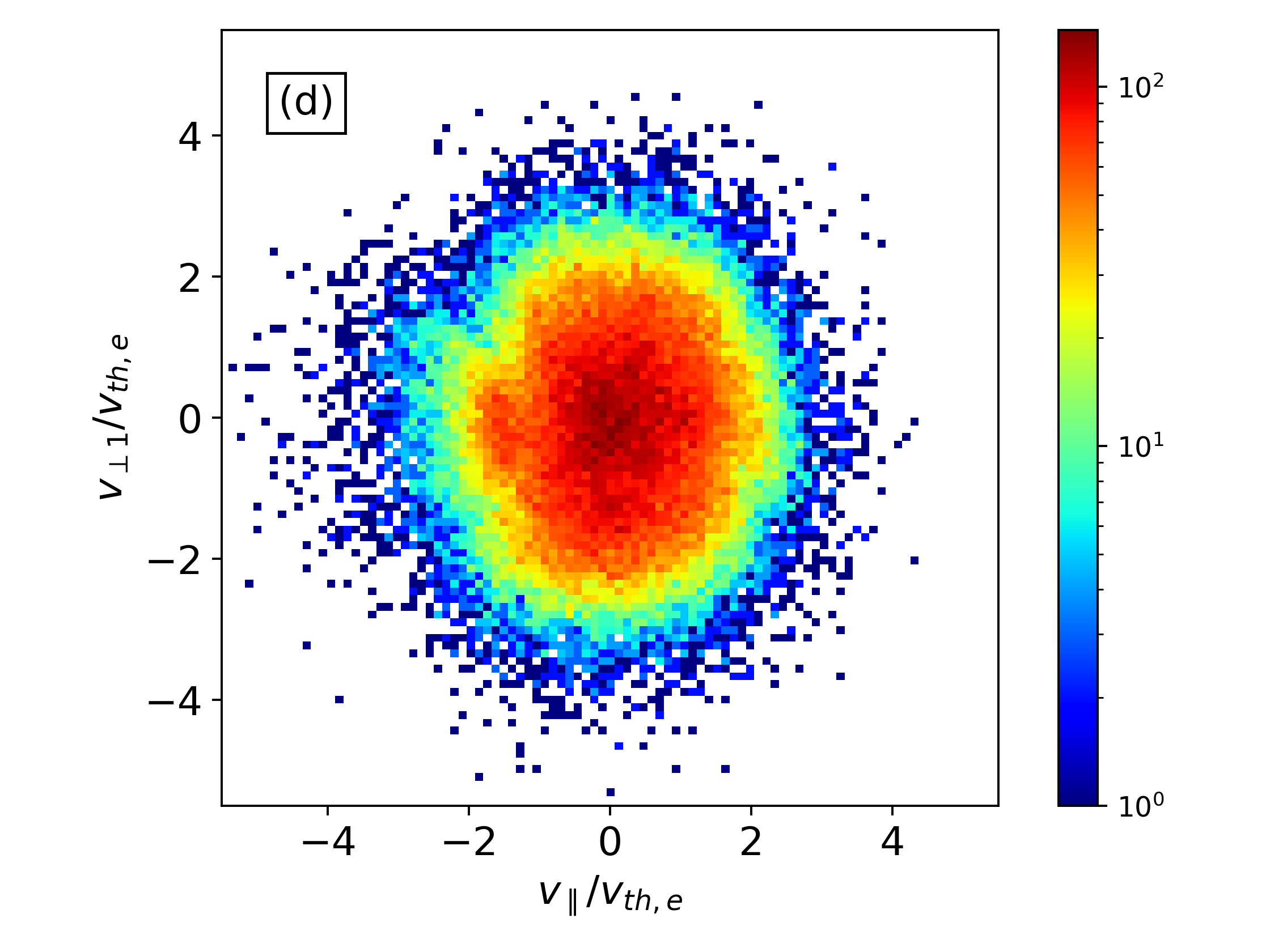} 
}
\hfill
\subfloat{
\includegraphics[width=0.32\linewidth]{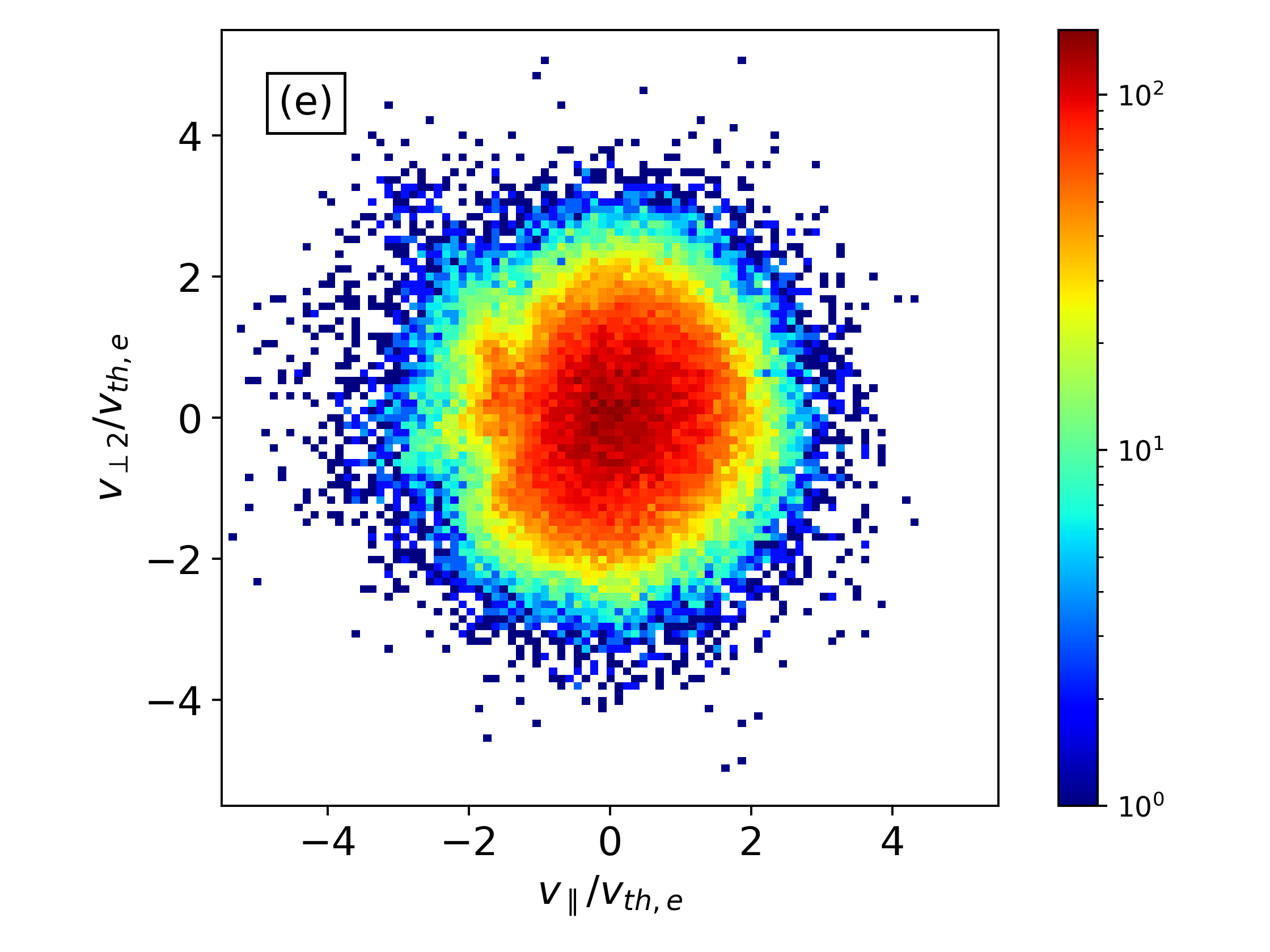} 
}
\hfill
\subfloat{
\includegraphics[width=0.32\linewidth]{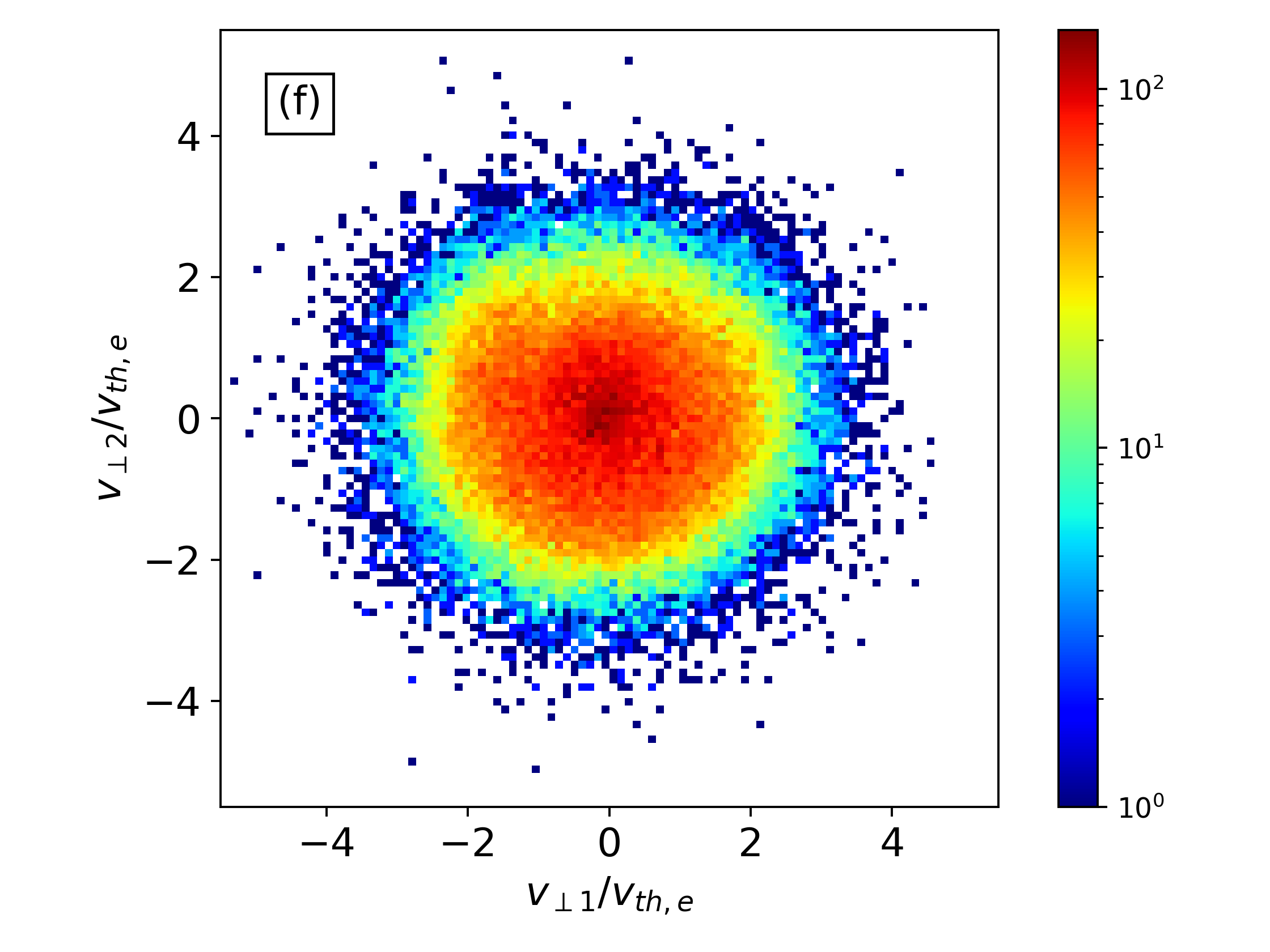} 
}
\end{minipage}
\caption{First example of EVMH at $t\!=\!650\,\Omega_e^{-1}$. (a) Shaded isocontours represent the magnetic field magnitude (in units of $B_0$) while black lines indicate the in-plane magnetic field lines. (b) Shaded isocontours represent the electron temperature anisotropy $T_{e,\perp}/T_{e,\parallel}-1$, black streamlines indicate the in-plane electron fluid velocity $\textbf{u}^{\prime}_e$ while the yellow square box highlights the region where the EVDF is calculated. (c) Three-dimensional isosurfaces of the EVDF, with the red arrow indicating the direction of the local magnetic field. (d)-(f) Two-dimensional projections of the EVDF in the directions parallel and perpendicular to the local magnetic field.}
\label{EVMH01}
\end{figure*}

In this section we discuss the properties of the EVMHs generated by the mechanism discussed in the previous section, i.e. by the combination of the large scale nonlinear dynamics and the sub-ion scale EKHI. After their formation, these structures persist also when the turbulence is fully developed and they appear to be stable, surviving for at least $100\,\Omega_e^{-1}$, unless they interact with other turbulent structures. Electric field fluctuations are negligible in correspondence of the EVMHs we observe and ion quantities do not show any significant correlation with these structures, consistently with the fact that ions are expected to be decoupled from the magnetic field dynamics at such small scales. For these reasons, we investigate only the magnetic field and electron quantities characterizing the EVMHs, focusing on the analysis of the electron velocity distribution functions (EVDFs). By comparing the EVDFs of all the EVMHs generated by the EKHI, we identify three main classes of distributions characterized by different shapes and features. Therefore, in this section we show and discuss three most representative examples of EVMHs, each of which is associated with one of these three classes of EVDFs (the other EVMHs observed in the simulation have EVDFs with shapes similar to the three we are going to show, with only minor differences). 

The first example of EVMH we analyze is the one produced by the EKHI event discussed in the previous section, visible in panel (f) of Figure~\ref{Formation} at position $(x, y)\!\simeq\!(8.1\,d_i, 28\,d_i)$. Figure~\ref{EVMH01} shows a close-up of this EVMH at $t\!=\!650\,\Omega_e^{-1}$. The shaded isocontours in panel (a) represent the magnetic field magnitude while black lines indicate the in-plane magnetic field lines. The EVMH appears as an oval-shaped region where the magnetic field amplitude suddenly drops, crossed by a bundle of quasi-parallel in-plane magnetic field lines that bend and swell up in correspondence of the structure, resulting in a magnetic field configuration typical of a mirror-like structure. This EVMH has an average diameter of the order of $d_i$ and smaller than the average $\rho_i$ which is equal to $\rho_i\!\simeq\!3.7\,d_i$ at fully developed turbulence. Panel (b) of Figure~\ref{EVMH01} shows the electron temperature anisotropy $T_{e,\perp}/T_{e,\parallel}-1$ in correspondence of the EVMH, where $T_{e,\perp}$ and $T_{e,\parallel}$ are the electron temperatures perpendicular and parallel to the local magnetic field, respectively. As observed in the figure, $T_{e,\perp}$ becomes larger than $T_{e,\parallel}$ inside the EVMH, which is a typical signature of this kind of structures, consistent with both satellite observations and previous numerical studies. Black streamlines indicate the in-plane electron fluid velocity $\textbf{u}^{\prime}_e$ which highlights the presence of the electron vortex associated with the magnetic field depression. To investigate the kinetic properties of the electrons, we consider an electron scale region inside the EVMH where we sample particles to calculate the EVDF. This region is indicated in panel (b) of Figure~\ref{EVMH01} by a yellow square box of size $1.25\,d_e$ and contains more than $90000$ electron macroparticles. Panel (c) of Figure~\ref{EVMH01} shows a 3D visualization of the EVDF thus obtained, with shaded isosurfaces corresponding to different shells where the distribution has a constant value. The EVDF is normalized to the number of macroparticles, so that its integral over the velocity space gives the total number of macroparticles inside the yellow box. The electron velocities are measured in units of the initial electron thermal speed $v_{th,e}$. The red solid arrow points in the direction of the mean magnetic field inside the yellow box where the EVDF is calculated. As observed in panel (c), the EVDF inside the EVMH is far from being Maxwellian and presents two main populations: an hot central core containing most of the particles and a colder, less populated beam. Both the core and the beam are almost aligned with the direction of the local magnetic field although the beam is partially off-center, making the EVDF slightly agyrotropic. The two populations are not isotropic, having larger velocities in the plane perpendicular to the local magnetic field, which implies a larger perpendicular temperature with respect to the parallel one, as consistent with the behavior of $T_{e,\perp}/T_{e,\parallel}-1$  shown in panel (b) of Figure~\ref{EVMH01}. The combination of these two electron populations results in a mushroom-shaped EVDF, with the hot anisotropic core representing the cap while the beam acts as the stem. These features can be observed also in the 2D projections of the EVDF in the directions parallel and perpendicular to the local magnetic field, shown in panels (d), (e) and (f) of Figure~\ref{EVMH01}. The magnetic field aligned parallel and perpendicular directions are defined as in \citet{goldman2016can}, using the three unit vectors $\hat{\textbf{e}}_{\parallel}$, $\hat{\textbf{e}}_{\perp 1}$ and $\hat{\textbf{e}}_{\perp 2}$ pointing in the directions of $\textbf{B}$, $\textbf{B}\times\hat{\textbf{z}}$ and $\textbf{B}\times\hat{\textbf{e}}_{\perp 1}$, respectively (where $\textbf{B}$ is the mean magnetic field in the region where the EVDF is calculated). 
The mushroom-like shape is roughly visible in panels (d) and (e) of Figure~\ref{EVMH01} where we show the two projections of the EVDF in the planes $(v_{\parallel},v_{\perp 1})$ and $(v_{\parallel},v_{\perp 2})$, respectively. The presence of the core and beam populations becomes less evident in 2D than in 3D as a consequence of the integration over the third velocity direction. However, it can still be observed in panel (e) that the beam is slightly tilted with respect to the parallel direction, a feature that implies a small agyrotropy on the EVDF, as already mentioned. Finally, panel (f) of Figure~\ref{EVMH01} shows the projection of the EVDF in the plane perpendicular to the magnetic field. The distribution looks quite isotropic and the presence of the tilted beam that makes the EVDF agyrotropic is less evident, appearing as a slight asymmetry in the direction of $\hat{\textbf{e}}_{\perp 2}$. This is again a consequence of the fact that the integration over the parallel velocity direction hides some features, especially those related to the beam which contains less electrons than the core and thus has a smaller weight in the integration. Lastly, we would like to point out that the specific box we have used to calculate the EVDF was not chosen for any particular reason and that the main properties of the distribution are essentially the same over the whole EVMH region, with only minor differences from place to place. 

\begin{figure*}[t]
\centering
\begin{minipage}{0.32\linewidth}
\subfloat{
\includegraphics[width=\linewidth]{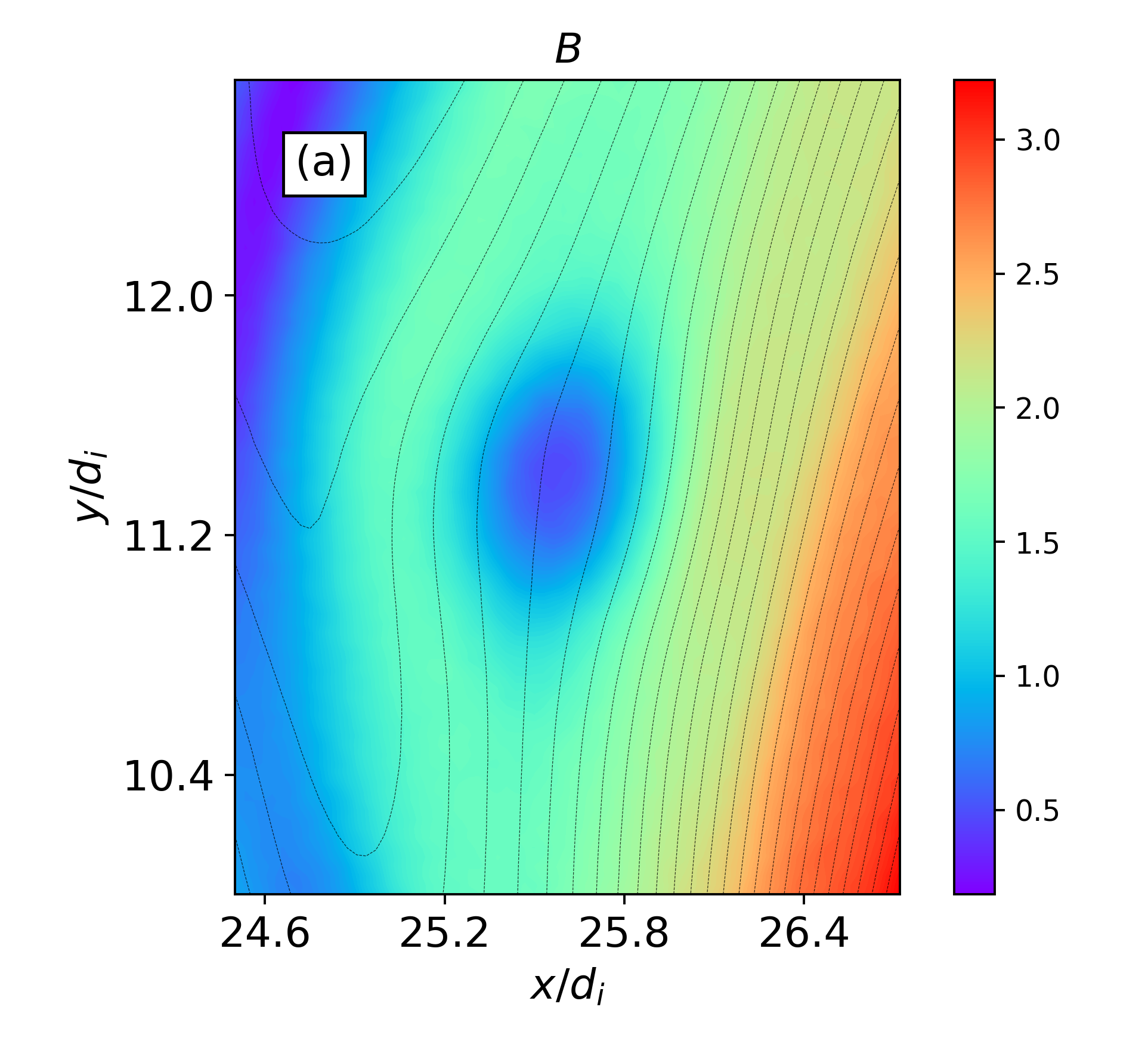} 
}
\vspace{-0.5cm}
\subfloat{
\includegraphics[width=\linewidth]{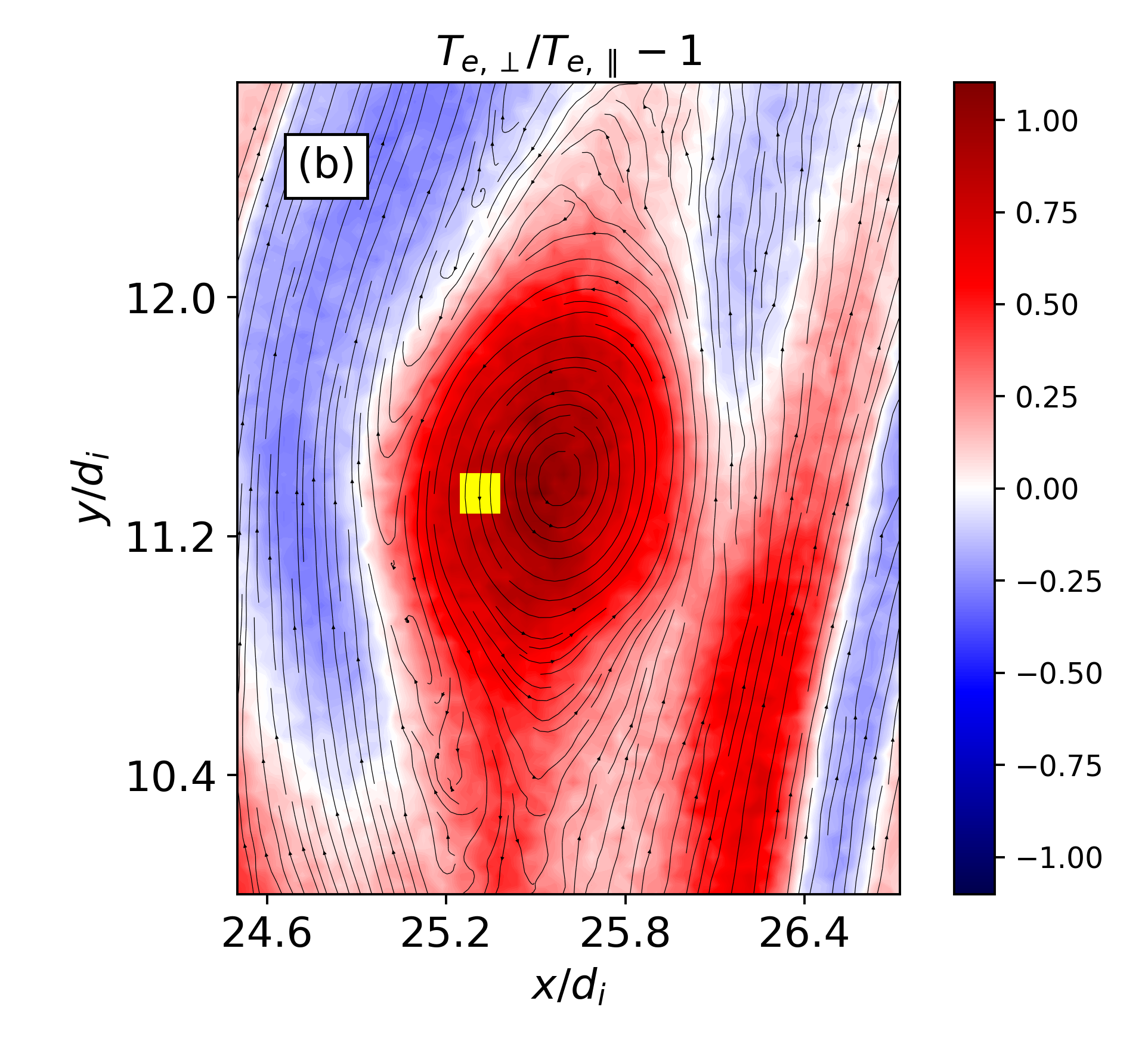} 
}
\end{minipage}
\hspace{0.9cm}
\begin{minipage}{0.58\linewidth}
\vspace{-0.5cm}
\subfloat{
\includegraphics[width=\linewidth]{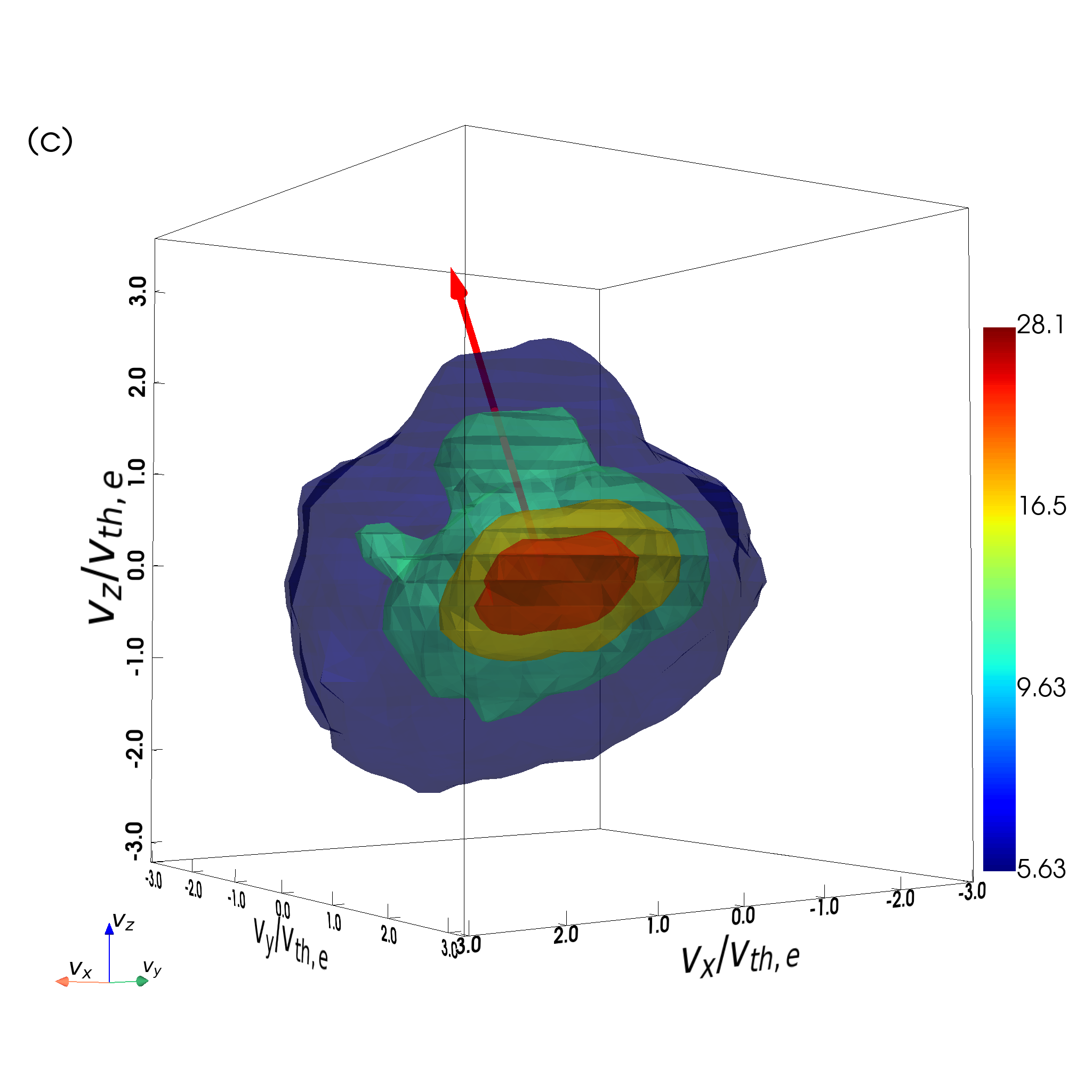}
}
\end{minipage}
\\
\begin{minipage}{\linewidth}
\subfloat{
\includegraphics[width=0.32\linewidth]{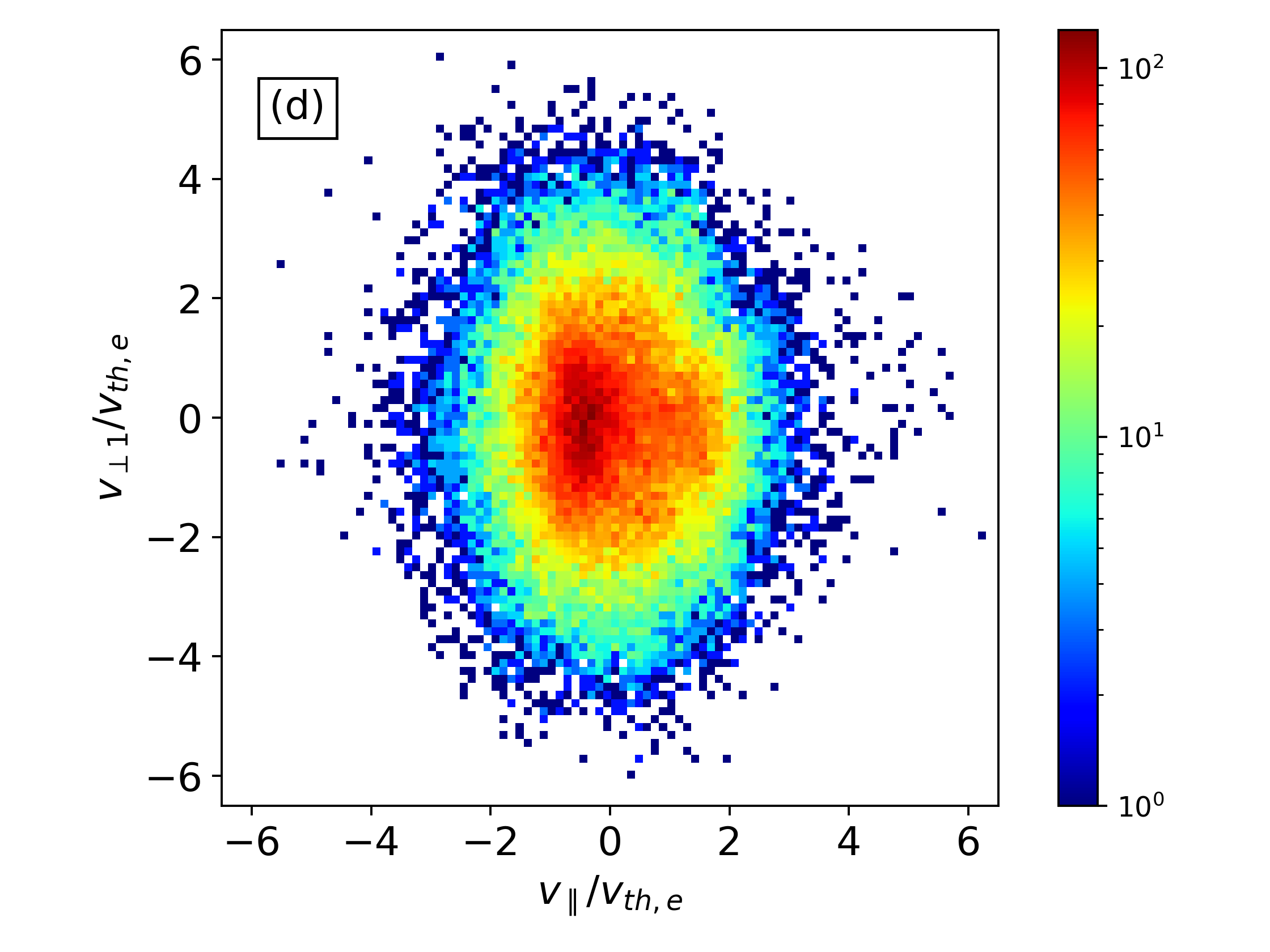} 
}
\hfill
\subfloat{
\includegraphics[width=0.32\linewidth]{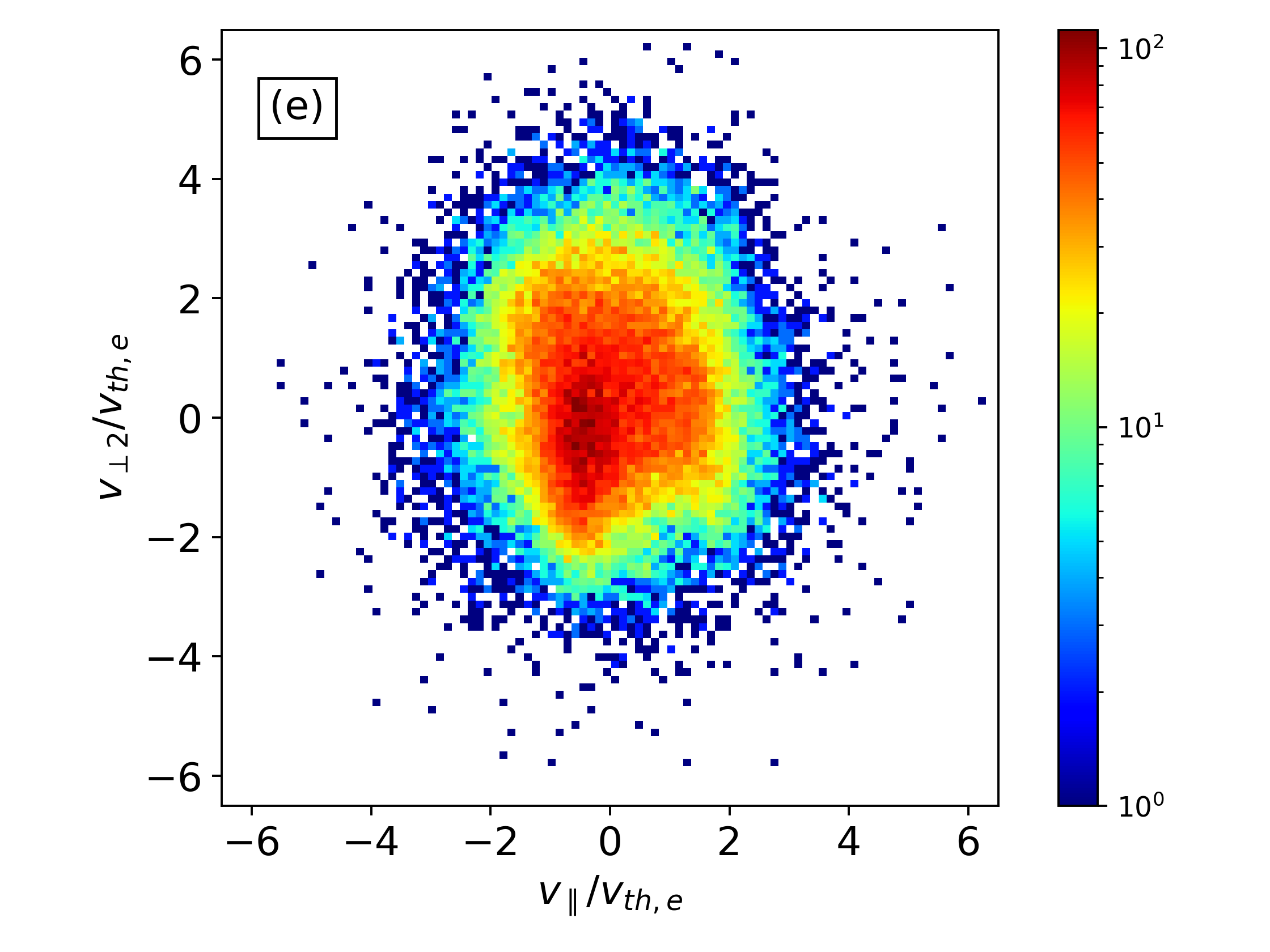} 
}
\hfill
\subfloat{
\includegraphics[width=0.32\linewidth]{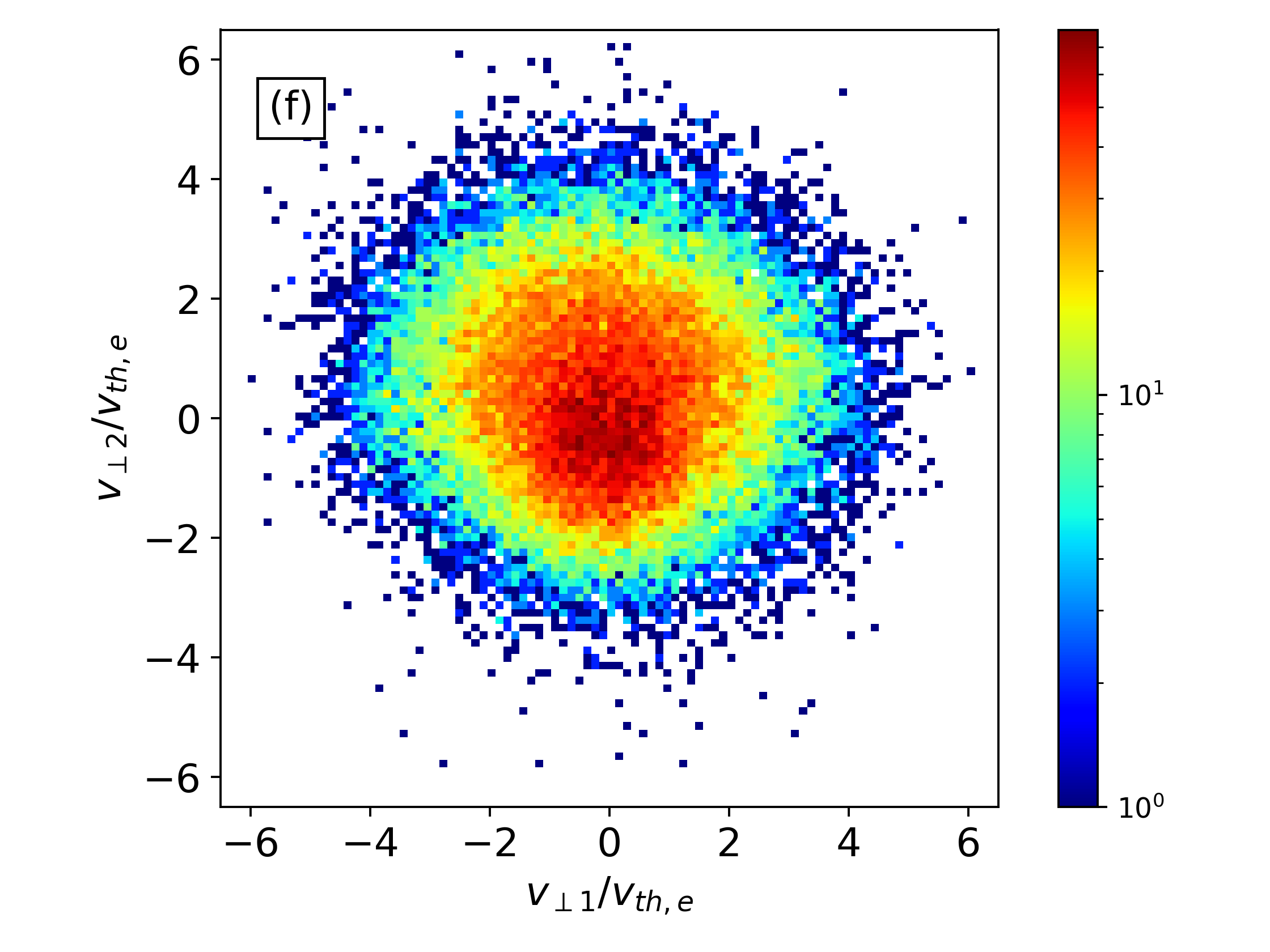} 
}
\end{minipage}
\caption{Second example of EVMH at $t\!=\!750\,\Omega_e^{-1}$. All the panels and elements in the figure are organized as in Figure~\ref{EVMH01}.}
\label{EVMH02}
\end{figure*}

\begin{figure*}[t]
\centering
\subfloat{
\includegraphics[width=0.48\linewidth]{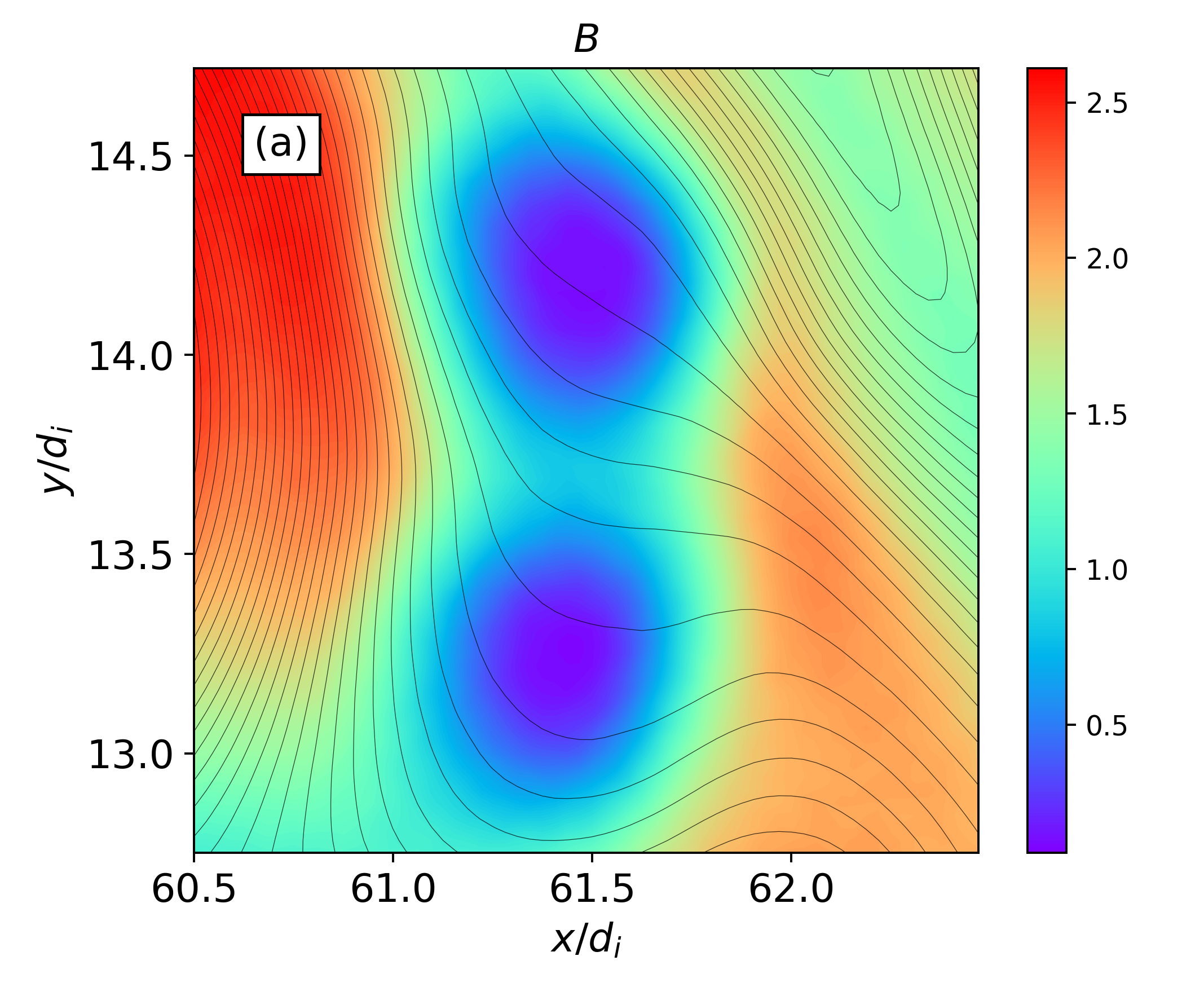}
}
\subfloat{
\includegraphics[width=0.48\linewidth]{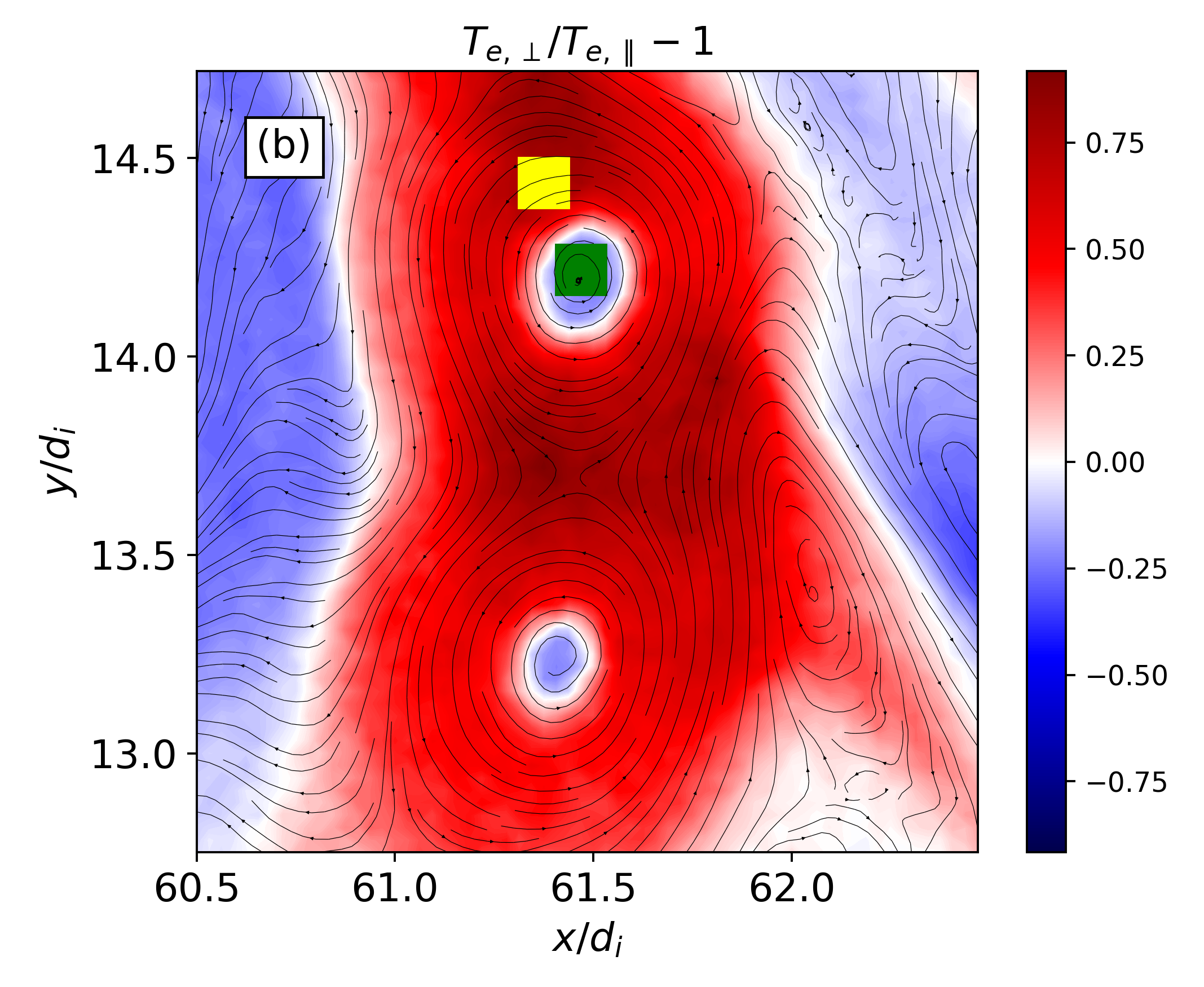}
}
\caption{Third example of a couple of EVMHs at $t\!=\!650\,\Omega_e^{-1}$. (a) Shaded isocontours represent the magnetic field magnitude (in units of $B_0$) while black lines indicate the in-plane magnetic field lines. (b) Shaded isocontours represent the electron temperature anisotropy $T_{e,\perp}/T_{e,\parallel}-1$, black streamlines indicate the in-plane electron fluid velocity $\textbf{u}^{\prime}_e$ while the yellow and green square boxes highlight the regions where the EVDFs are calculated.}
\label{EVMH03}
\end{figure*}

Figure~\ref{EVMH02} shows a close-up view at $t\!=\!750\,\Omega_e^{-1}$ of the second example of EVMH we analyze. All panels and elements in the figure are arranged in the same format as Figure~\ref{EVMH01}. By looking at panels (a) and (b) of Figure~\ref{EVMH02} we see that this second EVMH has several features in common with those observed for the first example we discussed. In particular, the structure has an average diameter of the order of $d_i$ and is characterized by a mirror-like magnetic field configuration. The electron temperature anisotropy $T_{e,\perp}/T_{e,\parallel}-1$ indicates that $T_{e,\perp}$ is larger than $T_{e,\parallel}$ in correspondence of the EVMH and the in-plane velocity $\textbf{u}_e^{\prime}$ is consistent with the presence of an electron vortex associated with the magnetic field dip. We investigate the kinetic properties of the structure by looking at the EVDF inside the EVMH, in the region indicated by the yellow square box of size $1.25\,d_e$ in panel (b) of Figure~\ref{EVMH02}. A 3D representation of the resulting EVDF is shown in panel (c) of the same figure and we note that its features are quite different with respect to the EVDF of the first EVMH we analyzed. In this case, the EVDF develops around a central core population and multiple beams are observed. The distribution presents a main prominent beam in the direction parallel to the local magnetic field (indicated by the red arrow) and is not isotropic, with larger velocities in the perpendicular plane. A couple of smaller beams are also present, not aligned with the local magnetic field (one of them is hidden behind the main beam in panel (c) of Figure~\ref{EVMH02}). Perpendicular velocities are distributed in the shape of a lobe, a feature that together with the smaller beams breaks the rotational symmetry of the EVDF around the magnetic field direction, making the distribution agyrotropic. This is another key difference with respect to the EVDF of the first EVMH we analyzed, where the agyrotropy resulted from the slightly tilted parallel beam alone rather than from the combination of multiple beams and a perpendicular lobe. Panels (d), (e) and (f) of Figure~\ref{EVMH02} show the 2D projections of the EVDF in the magnetic field aligned parallel and perpendicular directions. In panels (d) and (e) the anisotropy is clearly visible, with an evident larger velocity spread in the perpendicular direction than in the parallel direction. By comparing these projections with the corresponding ones in Figure~\ref{EVMH01}, we see that the different arrangement of the beams makes the temperature anisotropy of this second EVMH larger with respect to the first EVMH, as consistent with the values of $T_{e,\perp}/T_{e,\parallel}-1$ in correspondence of the two structures (see panels (b) of Figure~\ref{EVMH01} and Figure~\ref{EVMH02}). The beams are also visible in panels (d) and (e) of Figure~\ref{EVMH02}. In particular, the main beam is prominent in panel (d) while in panel (e) it gets mixed with the smaller beams because of the integration over the third velocity direction, resulting in a crest-shaped feature on top of the core of the distribution. This is another signature of agyrotropy since the two projections in panels (d) and (e) are not equivalent as it would have been in the case of a gyrotropic distribution. Panel (f) of Figure~\ref{EVMH02} highlights the presence of the fan-shaped lobe that additionally contributes in making the EVDF agyrotropic. We finally specify that also for this second example of EVMH, the main features of the EVDF are roughly the same over the entire structure, with some differences in the shape of the lobe that sometimes is less prominent, and in the intensity of the beams that in some regions contain either more or less electrons with respect to the sample we have shown. 

\begin{figure*}[t]
\centering
\subfloat{
\includegraphics[width=0.35\linewidth]{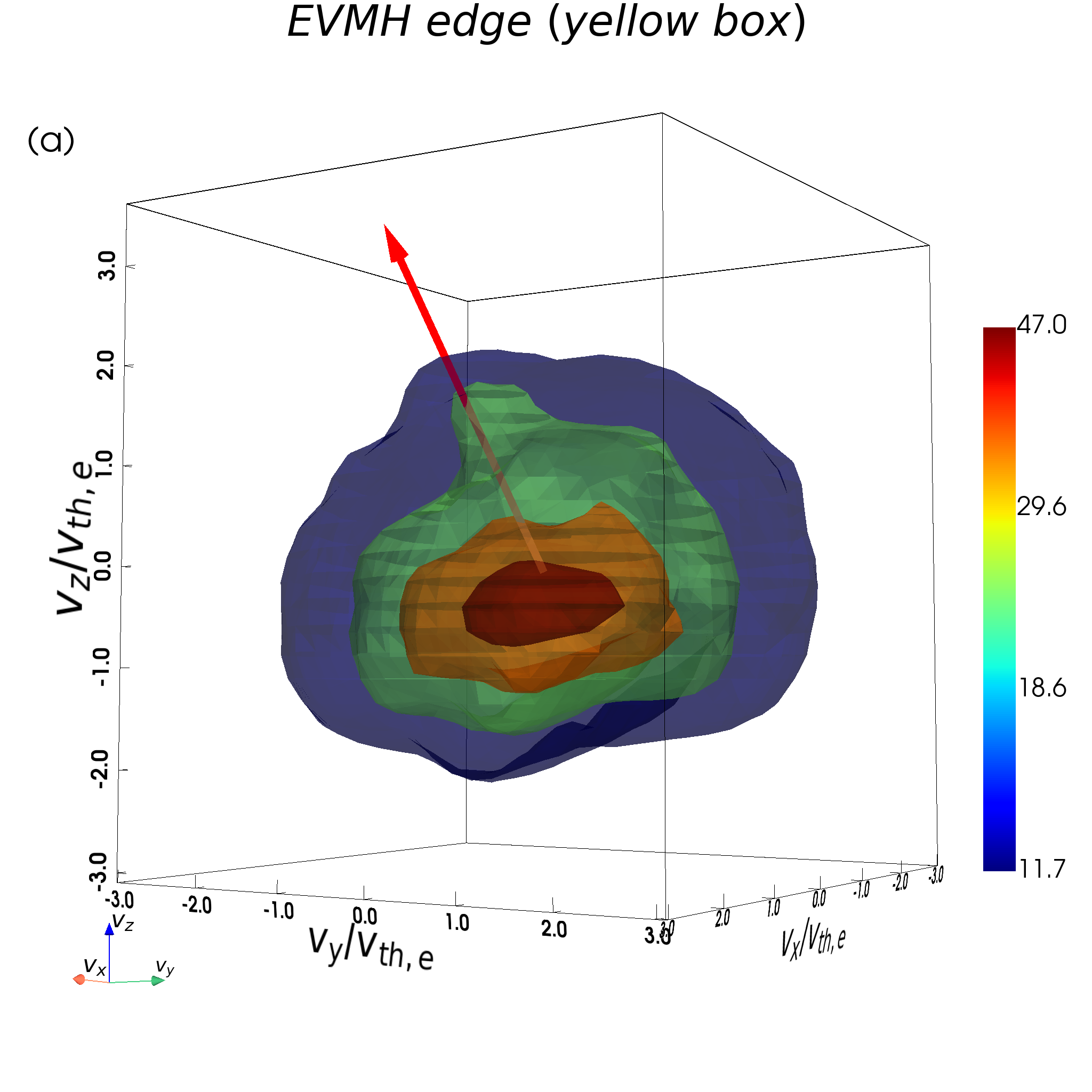} 
}
\hspace{0.9cm}
\subfloat{
\includegraphics[width=0.35\linewidth]{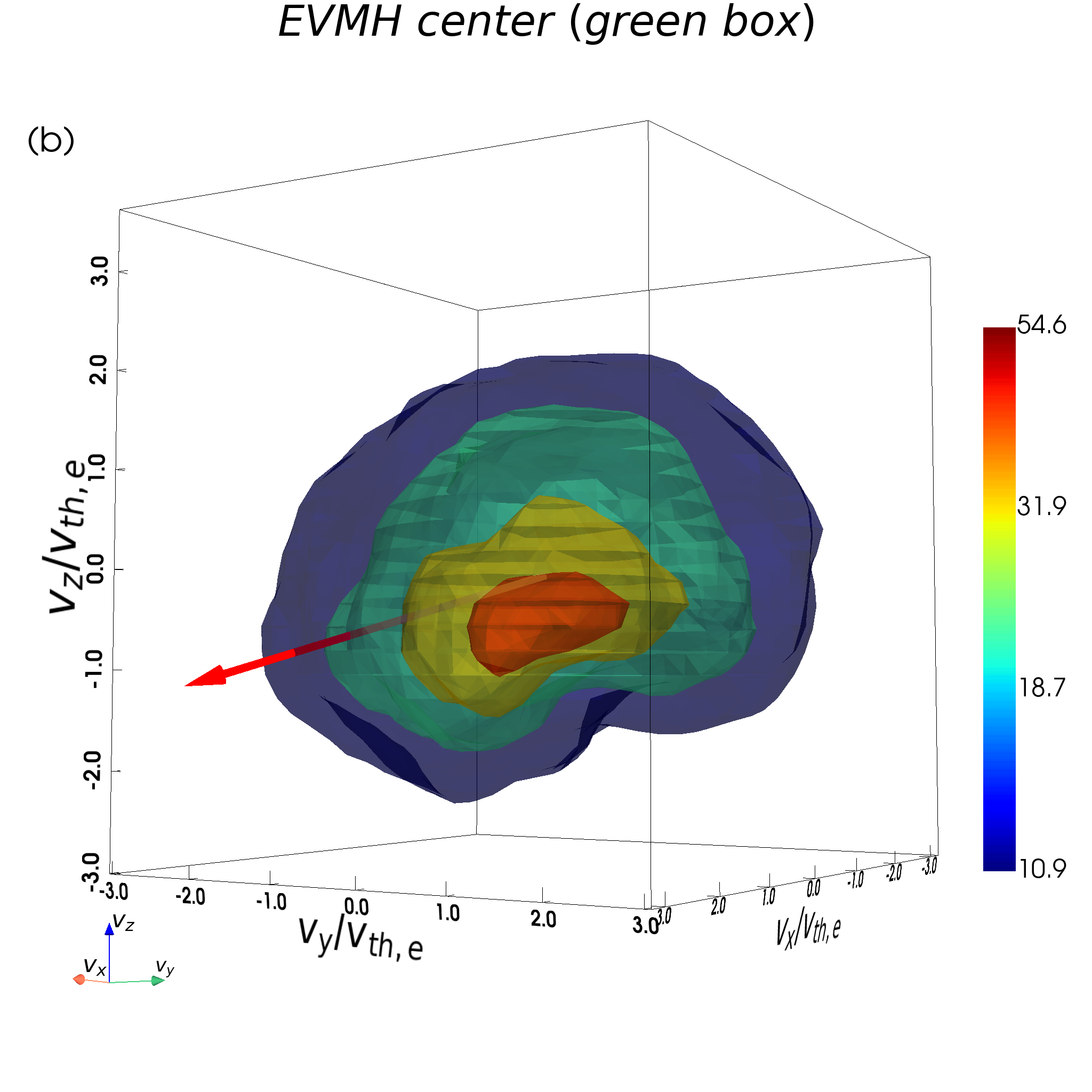} 
}
\vspace{-0.5cm}
\subfloat{
\includegraphics[width=0.8\linewidth]{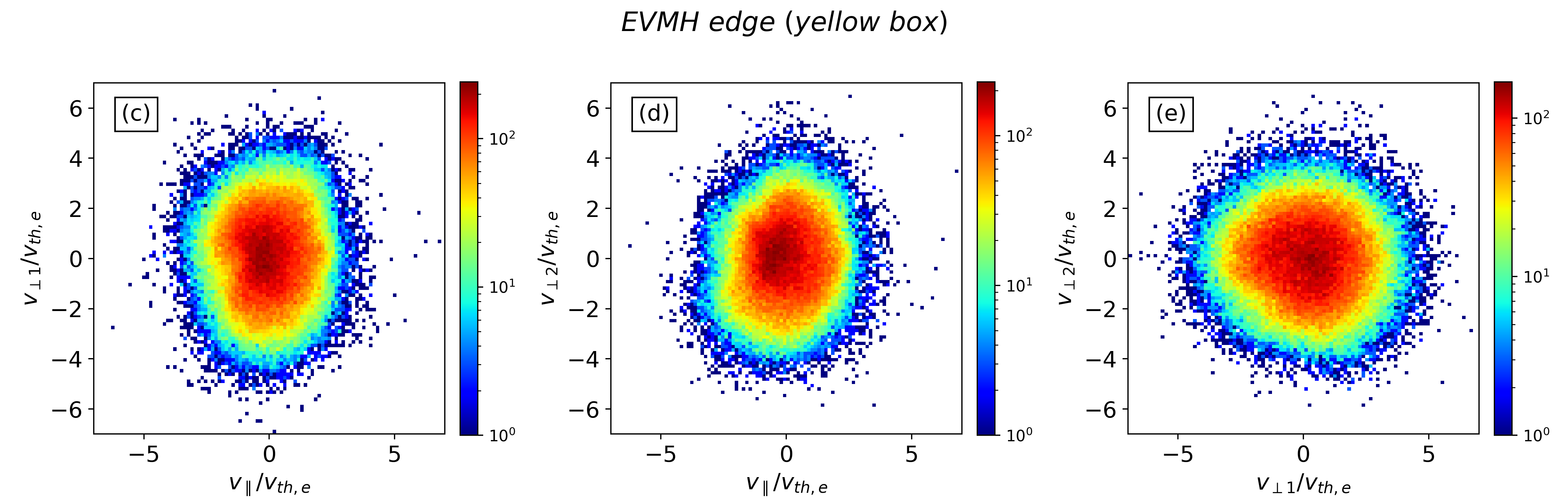}
}
\\
\subfloat{
\includegraphics[width=0.8\linewidth]{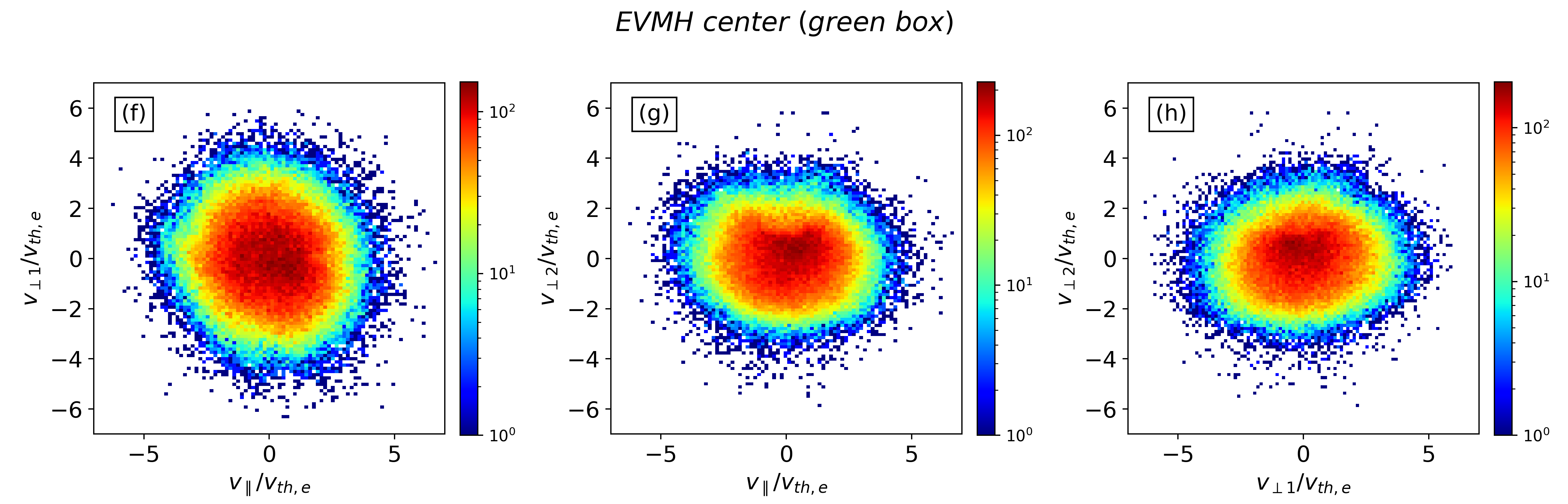} 
}
\caption{Three-dimensional isosurfaces of the EVDFs calculated on the edge (a) and in the center (b) of the EVMH of Figure~\ref{EVMH03}, with red arrows indicating the direction of the local magnetic field. Two-dimensional projections of the EVDF in the directions parallel and perpendicular to the local magnetic field, referred to the edge (c)-(e) and to the center (f)-(h) of the EVMH.}
\label{EVMH03_vdf}
\end{figure*}

The third and last example of EVMH we analyze is represented in Figure~\ref{EVMH03}, at $t\!=\!650\,\Omega_e^{-1}$. The figure shows a close-up of a couple of adjacent EVMHs generated by the same EKHI event. These structures exhibit unusual features with respect to the previous two examples. The shaded isocontours and black lines in panel (a) of Figure~\ref{EVMH03} represent the magnetic field magnitude and the in-plane magnetic field lines, respectively. These EVMHs are also characterized by an oval-shaped magnetic dip but the in-plane magnetic field deviates from a typical mirror-like configuration since magnetic field lines are significantly bent and deformed in the $-y$ direction as they cross the two EVMHs, instead of just swelling up as in the previous two cases. The presence of this magnetic field configuration is related to the fact that these EVMHs originated from an electron velocity shear that developed in a region where magnetic field lines had already been significantly bent because of the turbulent nonlinear dynamics. When the EKHI developed in this region, it only partially modified the in-plane magnetic field configuration that retained a signature of the initial magnetic tension, which was in turn inherited by the EVMHs after their formation. Another peculiar feature of these EVMHs can be observed in panel (b) of Figure~\ref{EVMH03} where the electron temperature anisotropy $T_{e,\perp}/T_{e,\parallel}-1$ is represented, together with the in-plane electron fluid velocity $\textbf{u}_e^{\prime}$, indicated by the black streamlines. We see that $T_{e,\perp}$ is larger than $T_{e,\parallel}$ everywhere inside the EVMHs, except for their center where the parallel electron temperature grows bigger than the perpendicular component. We investigate this change of behavior in the electron temperature anisotropy by analyzing the EVDFs calculated in two regions, one on the edge of the EVMH, where the anisotropy is positive, and the other one in the center of the structure, where the anisotropy is negative. These regions are indicated in panel (b) of Figure~\ref{EVMH03} by the yellow and green square boxes of size $1.25\, d_e$, placed on the edge and in the center of the EVMH, respectively. In this analysis we are only considering the upper EVMH, but analogous results are found also for the bottom one. The EVDFs obtained for the two regions are compared in Figure~\ref{EVMH03_vdf}. Panels (a) and (b) show a 3D visualization of the two distributions, with red arrows indicating the direction of the local magnetic field. We see that both EVDFs exhibit a similar dumpling-like shape with larger velocities in the $v_x$ and $v_y$ directions than along the $v_z$ direction, and a small concavity for $v_z\!<\!0$. A main core population is present in both cases and no significant differences are observed between the two EVDFs, except for a little beam in the $+v_z$ direction that is present in the distribution calculated on the edge (aligned with the local magnetic field) but not in the one calculated in the center. The change of sign in the electron temperature anisotropy when moving from the edge to the center of the EVMH can be understood by analyzing the relative orientation of the local magnetic field with respect to the corresponding EVDFs (the latter having essentially the same features in the two regions). From panel (a) of Figure~\ref{EVMH03_vdf} we see that on the edge of the EVMH the magnetic field is dominated by the $B_z$ component since it is mainly aligned with the $+v_z$ direction. This implies that on the edge of the EVMH the magnetic field points in a direction where the EVDF has a small velocity spread compared to the perpendicular direction, which means that $T_{e,\perp}\!>\!T_{e,\parallel}$ on the edge of the structure and thus the anisotropy is positive. On the other hand, from panel (b) of Figure~\ref{EVMH03_vdf} we see that $B_z$ drops abruptly in the center of the EVMH, where it becomes negative and smaller in amplitude than both $B_x$ and $B_y$, inducing a sudden change in the magnetic field direction which becomes almost perpendicular to $v_z$ (with a slight tilt in the $-v_z$ direction). As a consequence, the magnetic field in the center of the EVMH ends up aligning with a direction where the EVDF has a large velocity spread compared to the perpendicular direction, which means that $T_{e,\perp}\!<\!T_{e,\parallel}$ in the center of the structure and thus the anisotropy is negative. In other words, the change of sign in the electron temperature anisotropy between the edge and the center of the EVMH is not related to the presence of different features in the EVDFs in the two regions, but it simply results from a sudden change in the magnetic field orientation with respect to the distribution, induced by the strong $B_z$ reduction in the center of the structure. This effect can be observed also in the 2D projections of the EVDFs in the magnetic field aligned parallel and perpendicular directions. Panels (c), (d) and (e) of Figure~\ref{EVMH03_vdf} show these projections for the EVDF calculated on the edge of the EVMH. The dumpling-like shape is visible in panels (c) and (d), with an evident larger velocity spread in the perpendicular direction than in the parallel direction. The parallel beam is not really distinguishable, being hidden by the integration over the third velocity direction, while the small concavity is slightly visible for $v_{\parallel}\!<\!0$, especially in panel (c). Panel (e) shows that the EVDF on the edge of the EVMH has also some irregularities in the plane perpendicular to the local magnetic field, which make the distribution agyrotropic. By comparing these projections with those of the EVDF calculated in the center of the EVMH, represented in panels (f), (g) and (h) of Figure~\ref{EVMH03_vdf}, we observe that they are roughly equivalent by swapping $v_{\parallel}$ and $v_{\perp 2}$ of the distribution on the edge with $-v_{\perp 2}$ and $v_{\parallel}$ of the distribution in the center, respectively. This equivalence is indeed a consequence of the aforementioned sudden magnetic field rotation that takes place between the edge and the center of the EVMH because of the strong $B_z$ reduction. Therefore, as for the previous two examples of EVMHs we discussed, also for this unusual couple of EVMHs the EVDF does not show strong variations over the entire region occupied by the structures, and only minor differences are observed. Finally, we mention that there are other examples of EVMHs in our simulation exhibiting dumpling-shaped EVDFs similar to those of the two unusual EVMHs we have just discussed. Some of these EVDFs present a very prominent concavity for $v_{\parallel}\!<\!0$, more accentuated than in the case we showed, but they never show any sign change in the electron temperature anisotropy. Indeed, the two unusual EVMHs we showed as a third example are the only cases observed in our simulation where $T_{e,\perp}/T_{e,\parallel}-1$ changes sign when moving from their edge to their center. This is another evidence of the fact that the sign change in the anisotropy is not related to the specific shape of the EVDF and it just depends on the orientation of the magnetic field with respect to the distribution. As for the reason why this sign change in $T_{e,\perp}/T_{e,\parallel}-1$ is not observed in other EVMHs, we argue that it likely depends on the intensity of the in-plane current $\textbf{J}_{\perp}$ associated with the electron vortex supporting the magnetic depression. In fact, in order for $T_{e,\perp}/T_{e,\parallel}-1$ to change sign it is necessary to drastically modify the magnetic field direction by inducing a strong $B_z$ reduction between the edge and the center of the EVMH and this can be achieved only if the current in the associated electron vortex is intense enough. As already mentioned, the relation between $B_z$ and $\textbf{J}_{\perp}$ is a simple consequence of the Ampere's law $\nabla_{\perp} \!\times\! \left(B_z\,\hat{\textbf{z}}\right)\!\sim\!\textbf{J}_\perp$, which implies that only large $\textbf{J}_{\perp}$ can induce strong $B_z$ variations. Indeed, by comparing the intensity of the in-plane currents of the three examples of EVMHs we analyzed, we observe that the unusual one (the third example) exhibits a stronger $\textbf{J}_{\perp}$ which is about $1.9$ and $1.6$ times larger than those associated with the first and second examples, respectively. As a consequence, $B_z$ in the first two examples of EVMHs does not experience a drastic reduction, remaining positive and of the order of $B_x$ and $B_y$ even in the center of the structures.

As a final remark, another relevant point to mention is that although two of the EVHMs we showed have a mirror-like magnetic field configuration, the EVDFs inside these structures do not exhibit any evident loss cone feature which is typical of mirror-trapped particles. This happens because the size of the EVMHs is so small that the electron motion in correspondence of these structures is no longer adiabatic and does not conserve the magnetic moment, which implies that the electrons do not move on mirror-trapped orbits and the loss cone does not develop, as discussed in \citet{roytershteyn2015generation}. Indeed, in \citet{haynes2015electron} it has been shown that trapped electrons follow wide petal-shaped orbits inside EVMHs rather than performing the typical mirror-trapped motion where particles gyrate around magnetic field lines and are reflected back and forth at the edges of the mirror trap. In this sense, our results are consistent with and complementary to these previous studies. No loss cone signatures are found in any of the EVMHs generated by the EKHI in our simulation, even when these exhibit a mirror-like magnetic field configuration. This ultimately happens because the size of the EVMHs is of the order of the electron gyroradius, which implies that the electrons are only partially magnetized and their motion inside these structures consists of quite complex orbits rather than a simple gyromotion around magnetic field lines. Another implication of this partial demagnetization was observed in the third example of EVMH we discussed, where the magnetic field direction drastically changes on such a small scale that the electrons are not able to follow this quick variation. As a consequence, the EVDF maintains its shape and orientation across the whole structure, without adapting to the quick magnetic field rotation, which in turn causes the change of sign in the electron temperature anisotropy.

\section{Discussion and conclusions}

In this work we have studied the formation and the properties of the EVMHs developing in a fully kinetic 2D simulation of plasma turbulence initialized with parameters consistent with those observed in the Earth's magnetosheath.

We have identified a mechanism capable of generating sub-ion scale EVMHs from large scale fluctuations with wavelengths in the range $16\,d_i\!\leqslant\!\lambda\!\leqslant\!64\,d_i$. Such mechanism can be broken down as follows: first the nonlinear turbulent dynamics spontaneously produces thin and elongated electron velocity shears whose length can span up to about $10\,d_i$; as the turbulence further develops, the width of these electron velocity shears shrinks down to electron scales and they become unstable to the EKHI which tears them apart, producing sub-ion scale electron vortices; the electron current associated with these vortices reduces the local magnetic field and the resulting structures eventually evolve into EVMHs. This mechanism represents an effective way to channel the magnetic energy from large to sub-ion scales and our analysis highlights the significant role played by velocity shears and by the EKHI in the development of the turbulence and in the generation of sub-ion scale coherent structures. It is known that plasma turbulence naturally tends to generate magnetic field shears associated with intense current sheets \citep{servidio2009magnetic,servidio2010statistics}. The role of these structures in turbulence has been extensively studied in relation to the occurrence of energetic phenomena, such as reconnection, that can influence the properties of the turbulent cascade and of dissipation \citep{cerri2017reconnection,loureiro2017role,boldyrev2017magnetohydrodynamic,franci2017magnetic,dong2018role,comisso2018magnetohydrodynamic,arro2020statistical}. Here we have shown that electron velocity shears are also spontaneously generated by the turbulence and their disruption via the EKHI supports the cascade of energy from large to sub-ion scales, producing kinetic scale EVMHs that possibly contribute to dissipation. Indeed, it has been shown that MHs are associated with particles acceleration, heating \citep{liu2020electron} and due to their temperature anisotropy they can generate and host whistler waves and other kind of modes \citep{ahmadi2018generation,huang2018observations,yao2019waves}. Therefore, we argue that MHs may play a significant role in the reorganization of energy and dissipation in plasma turbulence, given their abundance in our simulation at fully developed turbulence. 

Regarding the implications of our work for space physics, we do not claim that the mechanism we discussed is the only one capable of generating sub-ion scale MHs but we argue that it may play a fundamental role in environments like the Earth's magnetosheath for instance, where other mechanisms are inhibited, such as the field-swelling instability that cannot develop since $T_i$ is typically larger than $T_e$, as in our simulation.

Another relevant point that needs to be mentioned is that this is not the first time that the link between the EKHI and the formation of sub-ion scale MHs is discussed. \citet{pritchett2009asymmetric} showed, using fully kinetic numerical simulations, that asymmetric guide field reconnection can generate small scale Kelvin-Helmholtz unstable electron velocity shears in the reconnection exhaust. The sub-ion scale current vortices produced by the instability can reduce the local magnetic field, generating a train of MHs. Recent satellite observations reported in \citet{zhong2022stacked} also show that the electron diffusion region in reconnection events is populated by kinetic scale electron vortices that are likely generated by the EKHI. Such vortices can locally enhance or reduce the magnetic field, producing magnetic peaks and holes depending on the direction of the electron vorticity with respect to the local magnetic field (i.e. depending on whether the vortex rotation is left handed or right handed with respect to the axis defined by the local magnetic field direction). In this context, our work shows that, beside magnetic reconnection, turbulence is another driver for the generation of sub-ion scale electron velocity shears whose instability can lead to the formation of kinetic scale MHs following a dynamics analogous to that discussed in \citet{pritchett2009asymmetric} and consistent with the EKHI.

\begin{figure*}[t]
\centering
\subfloat{
\includegraphics[width=0.48\linewidth]{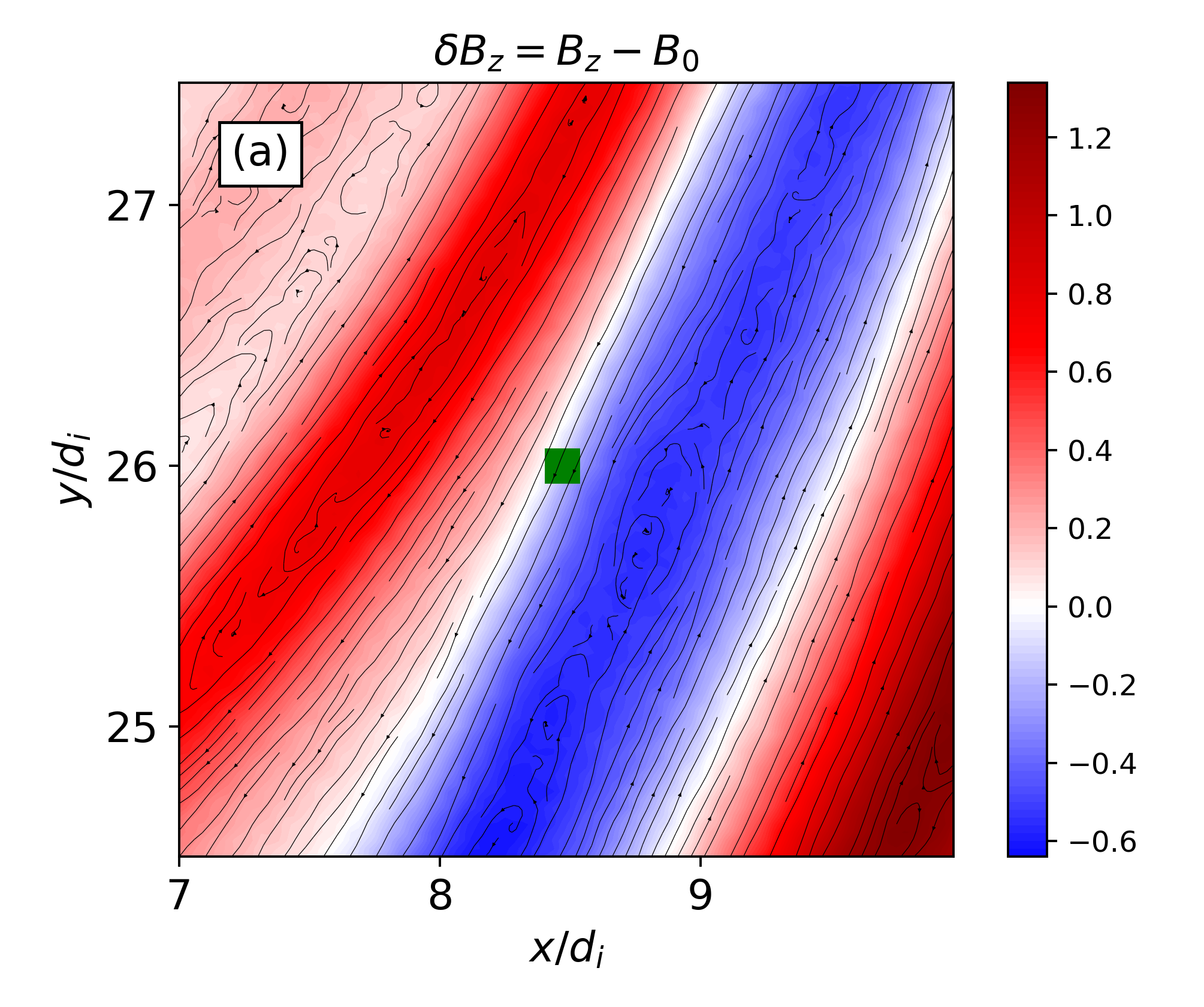}
}
\subfloat{
\includegraphics[width=0.48\linewidth]{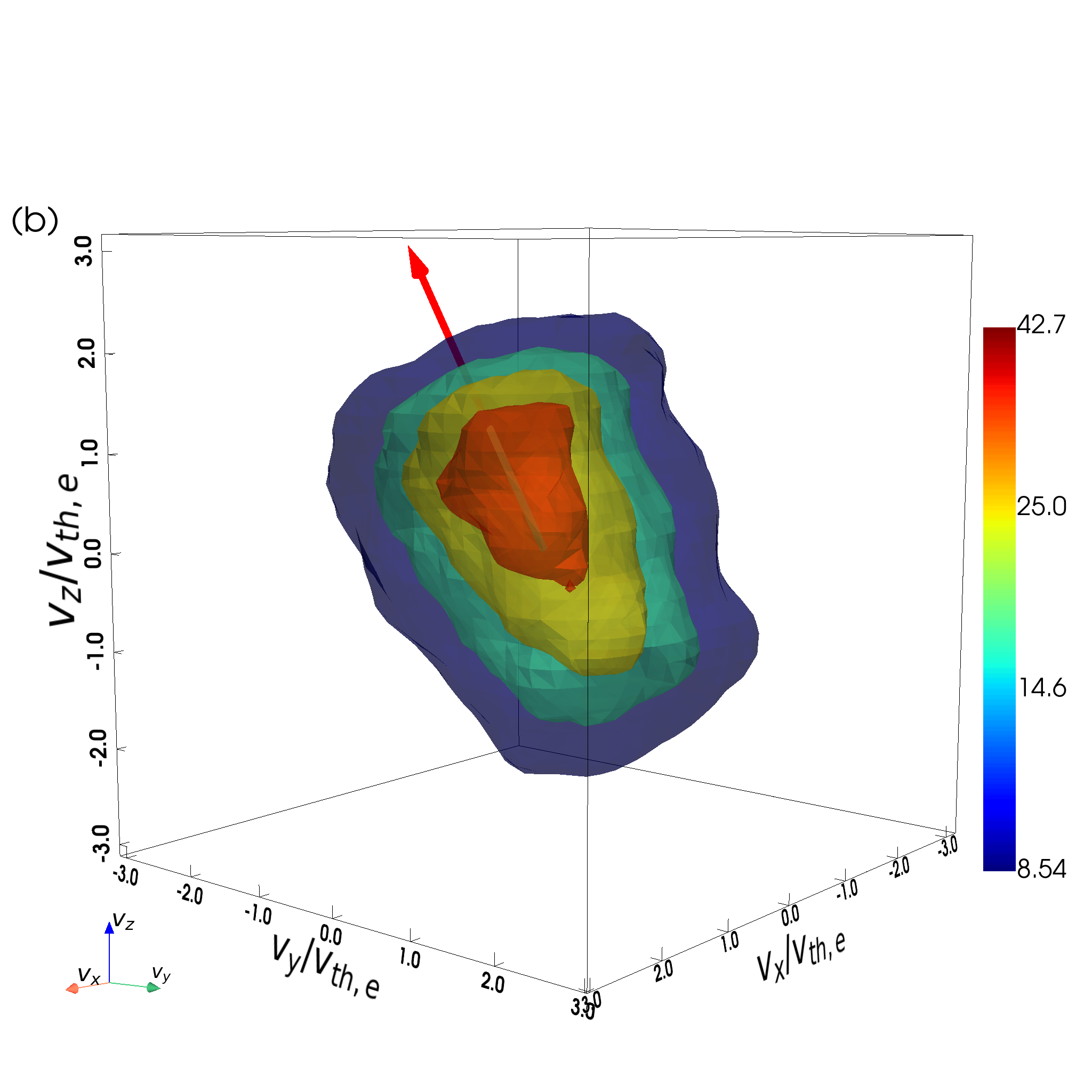}
}
\caption{(a) Close-up of the electron velocity shear at $t\!=\!350\,\Omega_e^{-1}$, where shaded isocontours indicate $\delta B_z$ (in units of $B_0$), black streamlines represent the in-plane electron fluid velocity $\textbf{u}^{\prime}_e$ while the green square box highlights the region where the EVDF is calculated. (b) Three-dimensional isosurfaces of the EVDF, with the red arrow indicating the direction of the local magnetic field.}
\label{shear}
\end{figure*}

We have investigated the kinetic properties of the EVMHs by analyzing their EVDFs and we identified three main classes of distributions, characterized by different shapes and features. The observed EVDFs are anisotropic, with a larger velocity spread in the plane perpendicular to the local magnetic field than in the parallel direction, and exhibit non-thermal features such as beams and other deformations that make them agyrotropic. Although all the EVMHs we analyzed are generated by the EKHI, it is not trivial to understand the reasons behind the differences in their EVDFs since their specific features are most likely determined by the details of the whole generation process, including the formation of the electron velocity shear, its disruption via the EKHI and the relaxation of the electron vortices into EVMHs. Besides, the dynamics that leads to the formation of these structures is intrinsically entangled with the turbulence and it is not easy to tell which features in the EVDFs are determined by the EKHI alone or by the turbulence. For instance, in \citet{haynes2015electron} it was shown that EVMHs (and magnetic depressions in general) can trap electrons via the grad-B drift and it cannot be excluded that some electron populations in the EVDFs we showed (for example the beams) may actually contain particles coming from neighbouring regions that have gotten trapped as they crossed the magnetic depression associated with either the electron velocity shear or with the EVMH. The interplay between coherent structures and turbulence is difficult to investigate since the turbulence is strongly nonlinear and all the structures immersed in it evolve simultaneously, possibly interacting and exchanging particles. A possible approach to get a better understanding of this complex dynamics would be to employ test particles simulations to track back the origin of the different populations observed in the EVDFs of the EVMHs. 

A complementary problem would be to study the EKHI in an isolated context, varying local plasma parameters and eventually adding an increasing level of turbulence to investigate how these different elements influence the generation of EVMHs and their properties. Our simulations shows that there is a certain degree of nonlinear coupling between the electron velocity shears after their formation and the underlying large scale turbulence. In fact, we mentioned that the shears undergo some slow modifications, like the steepening, and also their longitudinal structure, albeit stable, experiences slight large scale deformations (as for the case discussed in Figure~\ref{Formation} where the shear slightly bends over time). Therefore, a parametric study of the interplay between sub-ion scale electron velocity shears and turbulence is important to understand whether the EKHI alone is sufficient to produce EVMHs or if some turbulent forcing is indeed needed. The challenge here would be finding a suitable equilibrium configuration describing the electron velocity shear, which in a hot plasma at sub-ion scales requires either a fully kinetic treatment or more refined fluid models including finite Larmor radius corrections on the electron physics \citep{henri2013nonlinear,cerri2013extended,cerri2014pressure}. To our knowledge, so far the EKHI has only been studied in the context of cold EMHD using fluid equilibria where the structure of the velocity shear is maintained by the electric field alone that balances the Lorentz force \citep{jain2003nonlinear,jain2004kink,das2003sausage,gaur2009role,gaur2012linear}. However, in our simulation we see that the situation becomes way more complex in the case of an hot plasma and electron kinetic effects have to be considered to account for the presence of temperature anisotropies and heat fluxes associated with the velocity shear. As an example, in panel (a) of Figure~\ref{shear} we show a close-up of the same electron velocity shear considered in Figure~\ref{Formation}, at $t\!=\!350\,\Omega_e^{-1}$, before the EKHI develops. Shaded isocontours indicate $\delta B_z$ and the black streamlines represent the in-plane electron fluid velocity $\textbf{u}^{\prime}_e$. We analyze the EVDF calculated at the edge of this velocity shear, in the region indicated by the green square box of size $1.25\,d_e$. A 3D visualization of this EVDF is shown in panel (b) of the same figure and we see that the distribution is significantly non-Maxwellian. The red arrow indicates the direction of the local magnetic field and we observe an evident anisotropy with parallel velocities larger than the perpendicular ones. The distribution is also skewed in the direction of the magnetic field, which implies the presence of a parallel heat flux. All these features cannot be modelled by a simple fluid equilibrium and a fully kinetic description is necessary to properly study the EKHI at sub-ion scales with hot electrons.

Finally, the restriction to a 2D geometry represents a significant limitation in our simulation. It has been shown that EVMHs have a cocoon-shaped structure in 3D \citep{roytershteyn2015generation,wang2020three} whose origin and properties are still under debate. Therefore, 3D simulations are needed in order to investigate the formation of EVMHs in a more realistic environment and have results finally comparable with satellite observations. Nonetheless, 2D numerical simulations represent a necessary theoretical starting point to eventually interpret the 3D dynamics.


\begin{acknowledgments}

This project has received funding from the KULeuven Bijzonder Onderzoeksfonds (BOF) under the C1 project TRACESpace and from the European Union’s Horizon 2020 research and innovation program under grant agreement No. 955606 (DEEP-SEA).

This research was supported by the International Space Science Institute (ISSI) in Bern, through the ISSI International Team project 517: Towards a Unifying Model for Magnetic Depressions in Space Plasmas. 

The authors gratefully acknowledge the Gauss Centre for Supercomputing e.V. (www.gauss-centre.eu) for funding this project by providing computing time on the GCS Supercomputer SuperMUC-NG at Leibniz Supercomputing Centre (www.lrz.de) through the project “Heat flux regulation by  collisionless processes in heliospheric plasmas – ARIEL”. Additional computational resources have been provided by the Vlaams Supercomputer Centrum (VSC).

M.E.I. acknowledges support from the German Science Foundation (Deutsche Forschungsgemeinschaft, DFG) within the Collaborative Research Center SFB1491.

\end{acknowledgments}

\bibliography{EVMH}{}
\bibliographystyle{aasjournal}

\end{document}